\documentclass[12pt]{article}
\usepackage{cite}
\usepackage{graphicx}
\usepackage{epsfig}
\usepackage{amssymb}
\usepackage{rotating}
\usepackage{mathtools}
\usepackage{color,soul}

\def \gsim{\mathrel{\mathpalette\@versim>}}
\def \lsim{\mathrel{\mathpalette\@versim<}}
\def \neut{\tilde \chi^0}

\def \chargpm{\tilde \chi^\pm}
\def \glue {\tilde g}
\def \st {\tilde t}
\def \sq {\tilde q}
\def \sb {\tilde b}

\def \sign {\text{sign}}
\def \GeV {\text{ GeV}}
\def \TeV {\text{ TeV}}
\def \Tr {\text{Tr}}

\renewenvironment{thebibliography}[1]{%
\begin{oldthebibliography}{#1}%
\setlength{\parskip}{0ex}%
\setlength{\itemsep}{0ex}%
}%
{%
\end{oldthebibliography}%
}

\begin{document}

\vspace*{-1in}
\renewcommand{\thefootnote}{\fnsymbol{footnote}}

\vskip 5pt

\begin{center}
{\Large{\bf 
Compressed and Split Spectra\\[0.2cm]in Minimal SUSY SO(10)}}
\vskip 25pt

{
Frank F. Deppisch\footnote{E-mail: f.deppisch@ucl.ac.uk}$^1$,
Nishita Desai\footnote{E-mail: n.desai@thphys.uni-heidelberg.de}$^2$ and
Tom\'as E. Gonzalo\footnote{E-mail: tomas.gonzalo.11@ucl.ac.uk}$^1$
}

\vskip 10pt

{\it \small 
$^1$Department of Physics and Astronomy, University College London, UK \\
$^2$Institut f\"ur Theoretische Physik, Universit\"at Heidelberg, Germany 
}\\

\medskip

\begin{abstract}
\noindent
The non-observation of supersymmetric signatures in searches at the Large Hadron Collider strongly constrains minimal supersymmetric models like the CMSSM. We explore the consequences on the SUSY particle spectrum in a minimal SO(10) with large D-terms and non-universal gaugino masses at the GUT scale. This changes the sparticle spectrum in a testable way and for example can sufficiently split the coloured and non-coloured sectors. The splitting provided by use of the SO(10) D-terms can be exploited to obtain light first generation sleptons or third generation squarks, the latter corresponding to a compressed spectrum scenario.
\end{abstract}

\end{center}

\medskip

\renewcommand{\thefootnote}{\arabic{footnote}}
\setcounter{footnote}{0}

\section{Introduction}
\label{sec:introduction}
The non-observation of new heavy states at the LHC puts strong constraints on the sparticle spectrum of supersymmetric (SUSY) theories, especially in the coloured sector. Most importantly, this puts a strain on the ability of many SUSY models to solve the hierarchy problem of the Standard Model (SM) in a natural fashion. In minimal scenarios, such as the constrained minimal supersymmetric Standard Model (CMSSM), the stringent lower limits on coloured states will similarly affect non-coloured sparticles. The direct LHC search limits on these sparticle species as well as third generation squarks are on the other hand comparatively weak and can depend strongly on the details of the spectrum. Various solutions have been suggested to resolve the constraints and generate viable and testable scenarios. For example, phenomenological approaches like the phenomenological MSSM (pMSSM) do not contain a priori relations between different sparticle species and can be constructed to avoid the strong constraints but still provide states that can be produced at the LHC in the near future. On the other hand, such approaches often lack motivation.  

In this work, we focus on a minimal supersymmetric SO(10) model \cite{Fritzsch:1974nn,Anderson:1993fe,Raby:2003in} incorporating one-step symmetry breaking from SO(10) down to the Standard Model gauge group at the usual Grand Unified Theory (GUT) scale  $M_\text{GUT} \approx 2 \cdot 10^{16}$~GeV where the SM gauge couplings unify within an MSSM spectrum. Such a framework is therefore well motivated: It not only incorporates gauge unification but the unification of matter fields in a 16-plet would also provide degenerate soft SUSY breaking scalar masses at the GUT scale. In this scenario, the soft SUSY breaking sector is given by the gravity induced mass parameters for the matter and Higgs superfields at the GUT scale. Being a subset of the MSSM at low energies, two Higgs fields  are required to generate masses separately for up- and down-type fermions during electroweak symmetry breaking. In the SO(10) framework, these Higgs fields are generally produced from the superposition of doublet components in a set of Higgs fields at the GUT scale \cite{Aulakh:2007ir, Grimus:2006rk}. In the present analysis, we do not discuss the issue of Yukawa unification. Successful Yukawa unification of all fermion generations in SO(10) either requires a set of Higgs fields in large representations \cite{Aulakh:2006hs, Aulakh:2006vj, Aulakh:2007ir, Grimus:2006rk} or the presence of Planck-scale suppressed higher-dimensional operators \cite{Wiesenfeldt:2007us, Nath:2006ut}. 

In contrast to the CMSSM with its strictly degenerate soft scalar mass spectrum at the GUT scale, the scalar masses in the minimal SUSY SO(10) are non-universally shifted by D-terms associated with the breaking of SO(10) to the lower-rank SM group \cite{Drees:1986vd, Kolda:1995iw, Baer:2000gf}. These D-terms are analogous to the electroweak D-terms in the MSSM due to the rank reducing breaking of the SM gauge group. As described below in section~\ref{sec:model}, the SO(10) D-terms depend on the details of the breaking of SO(10) but are generally expected to be of the order of the SUSY breaking scale. They can therefore have a sizable impact on the sparticle spectrum. The possible presence of the SO(10) D-terms represents the main deviation from the CMSSM case, and we will analyze their impact on the sparticle spectrum in light of the LHC searches. As opposed to the phenomenological models, the non-degeneracy is not ad hoc and can be described by the introduction of a single additional parameter $m_D^2$. Starting at the GUT scale, the non-degenerate scalar masses evolve, following the renormalization group (RG) of the MSSM \cite{Jack:2003sx} down to the electroweak scale. This results in a sparticle spectrum at the supersymmetry scale chosen at 1~TeV according to the SPA convention \cite{AguilarSaavedra:2005pw}. If these masses were to be observed at the LHC or at other future colliders, the reverse RG evolution upwards would allow the reconstruction of the physics scenario at the GUT scale \cite{Freitas:2005et, Blair:2000gy, Blair:2002pg, Deppisch:2007xu, Deppisch:1900zz, Miller:2012vn, Miller:2013jra}.

In addition to the non-universality of scalar masses at the GUT scale due to SO(10) D-terms, we also allow for a non-degeneracy of the fermionic masses of the gauginos. While the gauge couplings unify at the GUT scale, the gauginos only do so if the messenger mediating the breaking of SUSY in a hidden sector is an SO(10) singlet \cite{Martin:2009ad}. This is not required though, and the messenger can be part of various SO(10) representations, provided it remains a singlet under the SM gauge groups.

This paper is organized as follows: In section~\ref{sec:model} we introduce the minimal SO(10) framework and the main consequences on the sparticle spectrum due to possible large D-terms and non-unification of the gaugino masses. Section~\ref{sec:direct_susy_searches} reviews the relevant direct sparticle mass limits from recent LHC searches. The results of our renormalization group analysis are presented in sectin~\ref{sec:Analysis} and we summarize our conclusions in section~\ref{sec:conclusions}.

\section{SUSY SO(10)}
\label{sec:model}
\begin{figure}[t]
\centering
\includegraphics[width=0.6\textwidth]{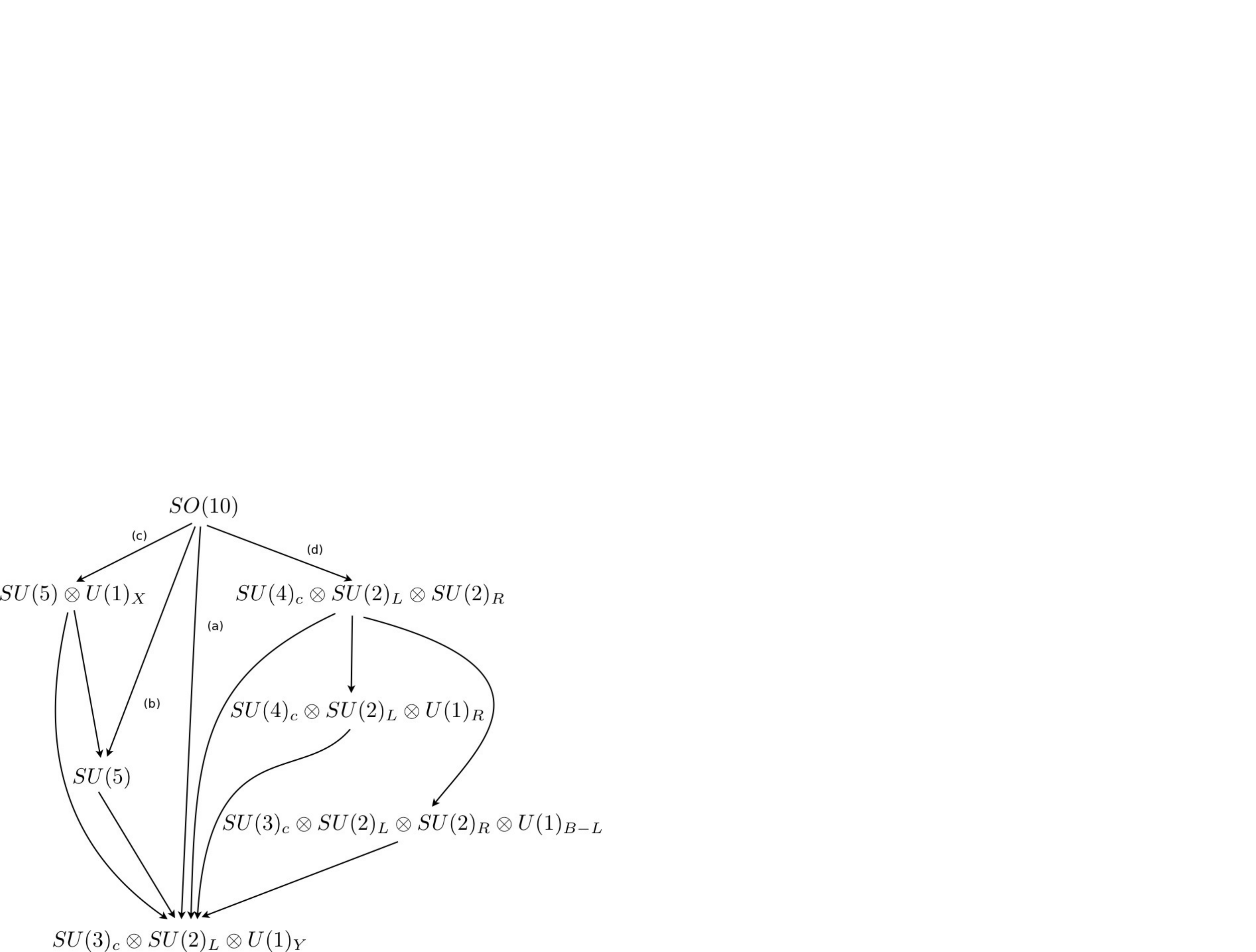}
\caption{Patterns of symmetry breaking from $\mathrm{SO}(10)$ to the Standard Model gauge group.}
\label{fig:SO10Breaking}
\end{figure}

SUSY GUT models are largely fixed by their gauge group structure. In SO(10), a generation of the SM fermions is contained in a $\mathbf{16}$ representation with the addition of a right-handed neutrino. Variations are then induced by the choice of the breaking of the GUT group to the SM group $\text{SU}(3)_c\times \text{SU}(2)_L\times \text{U}(1)_Y$. There are numerous ways in which this symmetry breaking can occur. A minimum of two breaking steps are required: one to break $\mathrm{SO}(10)$ to the SM group at a high scale $M_\text{GUT} \approx 2\times 10^{16}$~GeV (where the SM gauge couplings unify in the MSSM), and one to break the electroweak symmetry of the SM at $M_\text{EW}$. Among all the different possible breaking paths from $\mathrm{SO}(10)$ to $SU(3) \otimes SU(2) \otimes U(1)$, displayed in Figure~\ref{fig:SO10Breaking}, we will adopt the minimal path labeled (a). It should be noted that for phenomenological purposes, this is equivalent to multi-step breaking scenarios close to the scale $M_{GUT}$.

The electroweak Higgs fields of the MSSM are contained in higher-dimensional representations of $\mathrm{SO}(10)$, which couple to the SM fermions via Yukawa-type interactions. The only allowed representations for this field, given the SO(10) group structure, are $\mathbf{10}$, $\mathbf{120}$ and $\mathbf{126}$. We do not consider non-renormalizable operators which broaden the range of allowed Higgs representations. The simplest choice is to use the $\mathbf{10}$ dimensional representation containing the electroweak Higgs fields.  These choices motivate the superpotential
\begin{equation}
\label{Superpotential}
	W_\text{SO(10)} = \mathbf{Y} \mathbf{16}_F \mathbf{10}_H \mathbf{16}_F + \mu_H \mathbf{10}_H \mathbf{10}_H + W(\Sigma),
\end{equation}
where $\mathbf{Y}$ is a $3\times 3$ matrix in generation space. The term $W(\Sigma)$ collects all terms that involve the Higgs field(s) $\Sigma$ responsible for $\mathrm{SO}(10)$ breaking, which we can neglect in our low energy analysis.

The Higgs sector described above, i.e. the $\mathrm{SO}(10)$ breaking Higgs and the $\mathbf{10}_H$ containing the EW breaking Higgses, is not enough to predict the masses of all fermions in a Yukawa unified scenario. One would need to add larger representations and/or higher-dimensional operators, as mentioned before. However, extending this sector would not have a significant effect for the purpose of this study, for it is mostly focused on sfermion masses and any contribution coming from an extended Higgs sector can be neglected to the level of approximation at which we are working.

As phenomenologically required, SUSY has to be broken and the generated soft-SUSY breaking sector will depend on the particular breaking mediation mechanism. We assume Supergravity (SUGRA) mediated SUSY breaking where SUSY is broken above the GUT scale in a hidden particle sector. Before $\mathrm{SO}(10)$ breaking, these terms take the form
\begin{align}
\label{SoftTerms}
	\mathcal{L}_\text{soft} = 
	&- \mathbf{m}_{16_F}^2 \tilde{16}^*_F \tilde{16}_F 
	 - m_{10_H}^2 10_H^* 10_H \notag\\
	&- \frac{1}{2} m_{1/2} \tilde{X}\tilde{X}
	 -  A_0 \mathbf{Y} \tilde{16}_F \tilde{16}_F 10_H  
	 - B_0 \mu_H 10_H 10_H + c.c. \notag\\
	&+ \mathcal{L}_\Sigma,
\end{align}
where $\tilde{X}$ represents the gaugino field, $\tilde{16}_F$ and $10_H$ refer to the scalar components of the $\mathbf{16}_F$ and $\mathbf{10}_H$ superfields respectively. The corresponding soft breaking masses are denoted as $m_{1/2}$, $\mathbf{m}_{16_F}^2$ (in general a $3\times 3$ matrix in generation space) and $m_{10_H}^2$, respectively. The term $c.c.$ stands for complex conjugate and $\mathcal{L}_\Sigma$ collects any operators containing the $\Sigma$ field, which are irrelevant for our discussion. The SUSY breaking equivalents of the Yukawa coupling and Higgs $\mu$-term are controlled by the common trilinear coupling $A_0$ and $B_0$, respectively. In the following we will adopt the standard CMSSM boundary conditions for the trilinear soft-SUSY breaking parameters in the MSSM at the GUT scale:
\begin{equation}
	A_u = A_d = A_e = A_0.
\end{equation}
The corresponding boundary conditions for the soft scalar and gaugino masses will be discussed below.

\subsection{Scalar D-Terms}
\label{sec:Dterms}
The scalar potential of the $\mathrm{SO}(10)$ model, responsible for the symmetry breaking, is obtained from the scalar parts of the superpotential in  \eqref{Superpotential} plus the scalar soft breaking terms of  \eqref{SoftTerms}. In addition, there is an extra contribution that arises from the so called D-terms of the K\"ahler potential \cite{Kolda:1995iw}. Such D-terms are generated during gauge symmetry breaking that reduces the rank of the original group, i.e. when one or more of the embedded $U(1)$ subgroups is broken. The most prominent example is the electroweak D-term generated in the MSSM through the electroweak symmetry breaking (EWSB) of the SM gauge group to $\text{SU}(3)\times \text{U}(1)_Q$. For the breaking of a single U(1) subgroup, the process can be described as follows: The field acquiring a vacuum expectation value, $\Sigma$ in our case, has components with opposite charges under this U(1) subgroup, $\Phi$ and $\overline{\Phi}$ ($H_u$ and $H_d$ for EWSB). After symmetry breaking and after integrating out the heavy $\Phi$ and $\overline{\Phi}$, scalar particle masses receive contributions of the form \cite{Kolda:1995iw}
\begin{equation}
\label{DtermDependence}
	\Delta m_i^2 
	= Q_i m_D^2, \qquad \text{with} \qquad
	m_D^2 = \frac{1}{2}\frac{(\bar{m}^2 - m^2)}{ Q_{\Phi}},
\end{equation}
where $Q_i$ and $Q_\Phi$ are the charges of the light scalar particle species $i$ and the $\Phi$ field under the broken U(1), respectively. The soft masses of the $\Phi$ and $\overline{\Phi}$ fields are given by $m$ and $\bar{m}$, respectively, and they are related to the soft mass of the $\Sigma$ field(s) in \eqref{SoftTerms}. The D-term $m_D^2$ will therefore be roughly of the same order as the soft masses instead of the GUT scale where the breaking actually occurs. For more complicated breaking scenarios, the dependence of $m_D^2$ on the soft masses will vary slightly, according to the Higgs representation(s) involved, but it will still remain of the same order. In the case of EWSB, a linear combination of the $\text{U}(1)_Y$ and the U(1) included in $\text{SU}(2)_L$, generated by the $I_3$ generator, is broken. The electroweak D-terms has the value \cite{Drees:2004jm}
\begin{equation}
	\Delta m_i^2 = M_Z^2 \cos 2\beta(I^i_3 - Q_i \sin\theta_W),
	\label{EWDterm}
\end{equation}
with the third component of the weak isospin $I_3^i$ and the charge $Q_i$ of sparticle $i$ ($\tan\beta$ is the usual ratio of Higgs vacuum expectation values (VEVs)).

The contributions from the $\mathrm{SO}(10)$ D-term changes the boundary conditions for the scalar masses at the GUT scale. When the symmetry is spontaneously broken, the MSSM scalar masses match the $\mathrm{SO}(10)$ soft breaking masses in \eqref{SoftTerms}, plus the contributions from the D-term. Assuming that all soft-SUSY masses are diagonal and universal in generation space, the boundary conditions for the MSSM soft masses $\mathbf{m}_Q^2, \mathbf{m}_u^2, \mathbf{m}_e^2, \mathbf{m}_L^2, m_{H_d}^2, m_{H_u}^2$ read \cite{Aulakh:2007ir,Grimus:2006rk,Aulakh:2006hs,Aulakh:2006vj,Wiesenfeldt:2007us,Nath:2006ut,Drees:1986vd,Kolda:1995iw}
\begin{align}
\label{eq:BoundaryConditions}
	\mathbf{m}_Q^2 = \mathbf{m}_u^2 = \mathbf{m}_e^2 
	   &= m_{16_F}^2 \mathbf{1} + m_D^2\mathbf{1}, \notag \\
	\mathbf{m}_L^2 = \mathbf{m}_d^2 
	   &= m_{16_F}^2 \mathbf{1} - 3 m_D^2 \mathbf{1}, \notag \\
	\mathbf{m}_\nu^2 &= m_{16_F}^2 \mathbf{1} + 5 m_D^2 \mathbf{1}, \notag \\
	m_{H_d}^2 &= m_{10_H}^2 + 2m_D^2, \notag \\
	m_{H_u}^2 &= m_{10_H}^2 - 2m_D^2.
\end{align}
The coefficients in front of $m_D^2$ correspond to the $\mathrm{U}(1)$ charges of the different sparticles. This Abelian $\mathrm{U}(1)$ group is embedded into $\mathrm{SO}(10)$ via $\mathrm{SU}(5) \otimes \mathrm{U}(1) \subset \mathrm{SO}(10)$ and thus all particles in the same representation of $\mathrm{SU}(5)$ will have the same charge.
For completeness, we have also stated the boundary condition for the right-handed sneutrino soft mass $\mathbf{m}_\nu^2$. In the following, we will not consider the right-handed sneutrino as part of our spectrum. We implicitly assume it acquires a mass close to the GUT scale in a neutrino seesaw framework, and neglect the effect it could have on the running of the other sparticles as well as the lepton flavour violation it induces in the slepton sector. These effects depend delicately on the details of the neutrino sector. Equation~\eqref{eq:BoundaryConditions} describes the crucial impact of the presence of an SO(10) D-term. Most importantly it will cause a splitting between the sparticle species $\tilde Q, \tilde u, \tilde e$ and $\tilde L, \tilde d$ already at the GUT scale. This D-term induced splitting will be increased through RGE running, potentially causing a split spectrum at the low scales.

\begin{figure}[t]
\centering
\includegraphics[clip,width=0.7\textwidth]{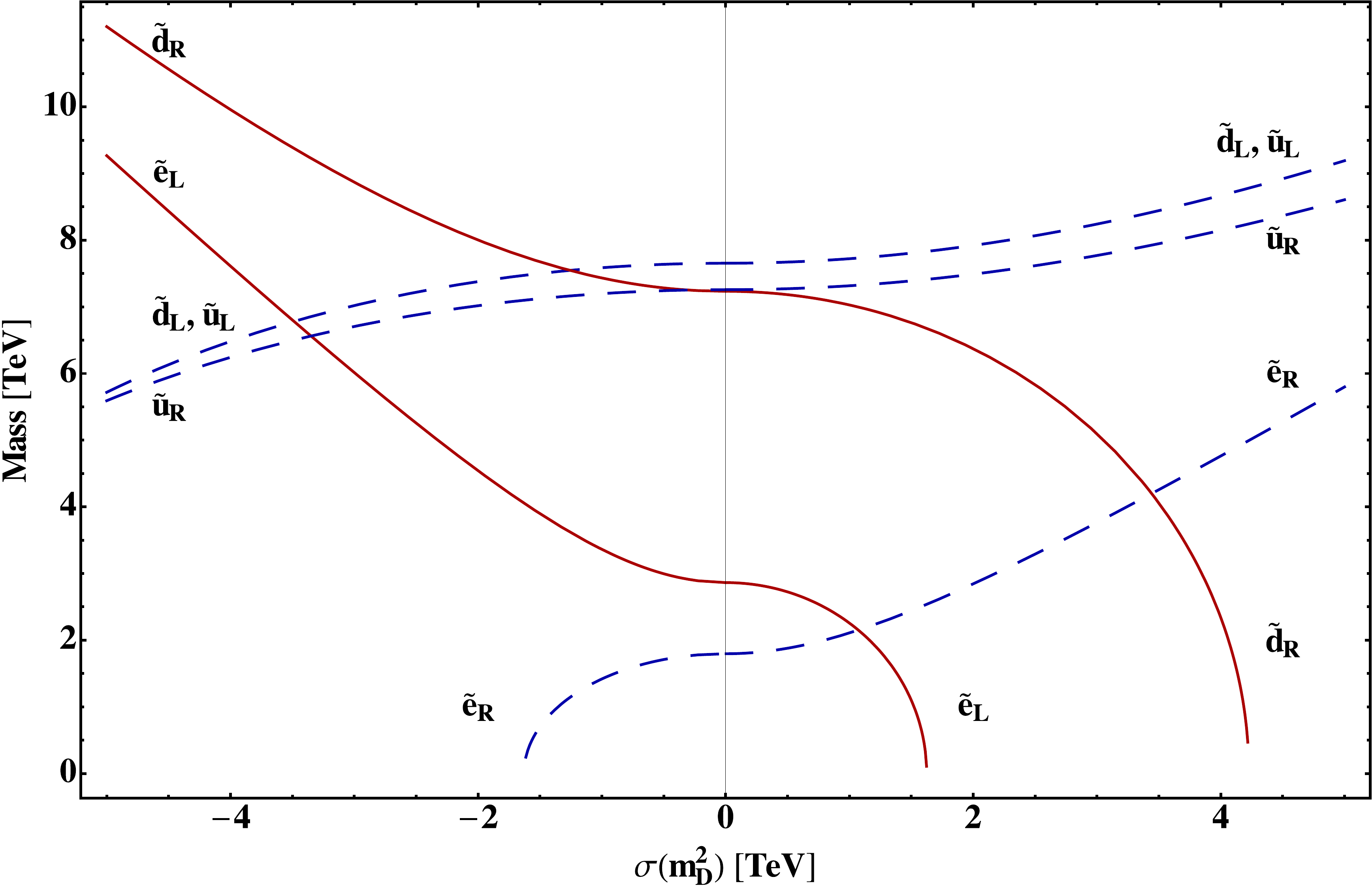}
\caption{First generation sfermion masses as a function of the SO(10) D-term $\sigma(m_D^2) = \sign(m_D^2) \sqrt{|m_D^2|}$. The values for the other model parameters are fixed as in Eq.~\eqref{BenchmarkScenario}.}
\label{fig:mD}
\end{figure}
The D-term will in general depend on the vacuum expectation value of the field that breaks the $\mathrm{SO}(10)$ gauge group, which in turn is related to the soft SUSY breaking masses as can be seen in the example \eqref{DtermDependence}. The specific value of the term depends very strongly on the scalar potential of the $\mathrm{SO}(10)$ breaking sector, but because we want to keep our description as independent as possible from the GUT scale physics, we will parametrize this by allowing $m_D$ to be a free parameter in our model. Thus, provided that the Yukawa couplings are fixed by the fermion masses up to the ratio of electroweak VEVs $\tan\beta = v_u/v_d$, and the $B_0$ and $\mu_H$ parameters are obtained by imposing electroweak vacuum stability conditions, the only free parameters of our model relevant to low energy phenomenology are
\begin{equation}
   m_{16_F}^2, m_{10_H}^2, m_D^2, m_{1/2}, A_0, \tan\beta, \sign(\mu_H).
\end{equation}
Figure \ref{fig:mD} shows how the masses of the first generation sfermions are split due the effect of the D-term. In order to present the dependence on the D-term $m_D^2$ in a convenient way, we define the function
\begin{equation}
   \sigma(m_D^2) = \sign(m_D^2) \sqrt{|m_D^2|}.
\end{equation}
The rest of the model parameters are fixed by using the benchmark scenario provided in Table 1 of \cite{Buchmueller:2013rsa},
\begin{gather}
  m_{16_F}     = 1380 \GeV, \quad 
  m_{10_H}     = 3647 i \GeV, \quad
  m_{1/2}      = 3420 \GeV, \notag\\ 
  A_0          = -3140 \GeV, \quad
  \tan\beta    = 39, \quad
  \sign(\mu_H) = 1,
\label{BenchmarkScenario}
\end{gather}
corresponding to a non-universal Higgs mass (NUHM1) high scale scenario.

%

\subsection{Non-Universal Gaugino Masses}
\label{sec:NonUniversalGauginos}
A standard assumption of the CMSSM is the unification of the gaugino masses at the GUT scale to the common value $m_{1/2}$ in \eqref{SoftTerms}. This is not necessarily true for more general SUSY breaking mechanisms. In particular, the $\mathrm{SO}(10)$ representation of the SUSY-breaking mediator field determines the matching conditions at the GUT scale. The field is required to be a singlet under the SM in order to preserve its symmetry but it does not need to be a singlet under $\mathrm{SO}(10)$. Table \ref{tab:GauginoBC} shows different boundary conditions for a selection of possible representations of the mediating field\cite{Martin:2009ad}. In the simplest case, the mediator field is in the singlet representation, in which case the matching conditions at the GUT scale are:
\begin{equation}
	M_1 = M_2 = M_3 = m_{1/2}.
\end{equation}
Other choices can have advantages, such as improved Yukawa unification\cite{Badziak:2012mm}. Other examples are models with negative $\mu_H$ which can be made compatible with the experimental value of the anomalous magnetic moment of the muon, $a_\mu$, by making $\mu M_2$ positive through the choice of a configuration with negative $M_2$ from Table~\ref{tab:GauginoBC}. 
\begin{table}[t!]
\centering
\begin{tabular}[c]{c|c|rr|rr}
& & \multicolumn{2}{|c|}{$M_\text{GUT}$} & \multicolumn{2}{c}{$M_\text{EW}$}  \\
\hline
SO(10) & SU(5) & $\tfrac{M_1}{M_3}$ & $\tfrac{M_2}{M_3}$ & $\tfrac{M_1}{M_3}$  & $\tfrac{M_2}{M_3}$ \\
\hline
$\mathbf{1}, \mathbf{54}, \mathbf{210}, \mathbf{770}$ & $\mathbf{1}$  & 
   1  & 1  & $\tfrac{1}{6}$ & $\tfrac{1}{3}$ \\
 $\mathbf{54}, \mathbf{210}, \mathbf{770}$ & $\mathbf{24}$ &  
   -$\tfrac{1}{2}$  & -$\tfrac{3}{2}$ & -$\tfrac{1}{12}$ & -$\tfrac{1}{2}$ \\
$\mathbf{210}, \mathbf{770}$ & $\mathbf{75}$ & -5 & 3 & -$\tfrac{5}{6}$ & 1\\
$\mathbf{770}$ & $\mathbf{200}$ & 10 & 2 & $\tfrac{5}{3}$ & $ \tfrac{2}{3} $ \\
\hline
\end{tabular}
\caption{Ratios of gaugino masses for a SUSY breaking messenger field in different representations of $\text{SU}(5) \subset \text{SO}(10)$ \cite{Martin:2009ad} at the GUT and the EW scale. The EW ratios take into account the approximate effect of the RGE running on the gaugino masses.}
\label{tab:GauginoBC}
\end{table}

In models that undergo gauge mediated supersymmetry breaking, this non-universali\-ty emerges naturally at the messenger scale due to the nature of the breaking. At this messenger scale, usually around or above $10^{6} \GeV$, the masses of gauginos are induced by one-loop corrections involving messenger fields, and are of the form \cite{Dine:1993yw}
\begin{equation}
   M_a = \frac{\alpha_a}{4\pi} \Lambda \sum_{n_a} n_a,
\end{equation}
where $\Lambda$ is the relative splitting of the fermionic and scalar parts of the messenger superfields (source of supersymmetry breaking) and $n_a$ is the Dynkin index of the messenger fields in the SM subgroup $a$. In this case there can be two sources of non-universality: first, there is a natural splitting due to the different values of the gauge couplings $\alpha_a$, and second, the sum of the Dynkin indices could naively be different for the three gauge groups. However, if these messengers come in complete representations of the unified group (in order to preserve the unification of gauge couplings), the sum of the Dynkin indices is the same for all three gauginos. In this case, the only splitting at the messenger scale comes from the different values of $\alpha_a$, which can be rather small, and depends mostly on the messenger scale. In this paper we will focus only on mSUGRA-inspired scenarios, where the only non-universality in the gaugino mass comes from the $\mathrm{SO}(10)$ representation of the mediator field. Unless otherwise stated, we will consider universal gauginos at the GUT scale, with mSUGRA induced supersymmetry breaking. The effect of having non universal gauginos on the particle spectrum will be studied in section \ref{sec:NonUniversalGauginosAnalysis}.

\subsection{Renormalization Group Evolution}
\label{sec:RGEs}
Below the GUT scale, with the heavy gauge bosons and Higgs fields integrated out, the particle content of the minimal SUSY SO(10) model is the same as in the MSSM. We implicitly assume that the right-handed neutrinos and sneutrinos also decouple at or close to the GUT scale within a seesaw framework of light neutrino mass generation. Therefore the Renormalization Group Equations (RGEs) will be same as those of the MSSM but with different boundary conditions at the GUT scale. The complete RGEs for the MSSM and their approximate solutions are listed in Appendix \ref{app:RGEs}. In this section we will focus on the relevant consequences for the sparticle spectrum in the minimal SUSY SO(10) model using appropriate approximations. 

The RGEs for the scalar masses of the first two generations can be exactly solved at one loop by neglecting small Yukawa couplings. For the very same reason, there is no mixing between the left and right-handed squarks or sleptons under such an approximation. The RGEs are then given by
\begin{align}
	16\pi^2\frac{d}{dt}m_{Q_{1,2}}^2 &= - \frac{32}{3} g_3^2 M_3^2 - 6 g_2^2 M_2^2 - \frac{2}{15} g_1^2 M_1^2 + \frac{1}{5}g_1^2 S, \notag\\
	16\pi^2\frac{d}{dt}m_{u_{1,2}}^2 &= - \frac{32}{3}g_3^2M_3^2 -\frac{32}{15} g_1^2 M_1^2 - \frac{4}{5} g_1^2 S, \notag \\
	16\pi^2\frac{d}{dt}m_{d_{1,2}}^2 &=  - \frac{32}{3}g_3^2 M_3^2 - \frac{8}{15} g_1^2 M_1^2 + \frac{2}{3} g_1^2 S, \notag \\
	16\pi^2\frac{d}{dt}m_{L_{1,2}}^2 &= - 6g_2^2 M_2^2 - \frac{6}{5} g_1^2 M_1^2 - \frac{3}{5} g_1^2 S, \notag\\
	16\pi^2\frac{d}{dt}m_{e_{1,2}}^2 &= - \frac{24}{5} g_1^2 M_1^2 + \frac{6}{5} g_1 S,
\end{align}
with the gauge couplings $g_i$ and gaugino masses $M_i$. The term $S$ is defined as
\begin{equation}
	S = m^2_{H_u} - m^2_{H_d}  + \text{Tr} \left(m^2_Q - 2m^2_u + m^2_d - m^2_L + m^2_e \right).
\end{equation}
Although $S$ has a dependence on all the scalar masses, this particular combination turns out to be exactly solvable, and the solution depends only on the gauge couplings and the value of $S$ at the GUT scale. However, in the case that all scalar masses are universal, i.e.\ have the same value at the GUT scale, this term vanishes. It therefore has the role of quantifying the non-universality of a model. In our particular case, the universality is violated due to the appearance of the D-term, and so the only contribution left from this $S$ term is proportional to $m_D^2$. Thus the masses for all first and second generation squarks and sleptons can be expressed analytically as \cite{Miller:2012vn}
\begin{alignat}{8}
& m_{\tilde{u}_L}^2    =  \quad && m_{16_F}^2 +  \quad && m_D^2  \left(1 + 2C_1^{(1)}\right) 
									  && + \quad && m_{1/2}^2 \left(C_3^{(2)} + C_2^{(2)} + \frac{1}{6}C_1^{(2)}\right)  && + \quad && D_{u_L}, \notag \\
& m_{\tilde{u}_R}^2   = \quad && m_{16_F}^2 + \quad && m_D^2 \left(1 - 8C_1^{(1)}\right)
									  && + \quad && m_{1/2}^2 \left(C_3^{(2)} + \frac{8}{3}C_1^{(2)}\right)  && + \quad && D_{u_R}, \notag \\
& m_{\tilde{d}_L}^2    = \quad && m_{16_F}^2 + \quad && m_D^2 \left(1 + 2C_1^{(1)}\right)
									  && + \quad && m_{1/2}^2 \left(C_3^{(2)} + C_2^{(2)} + \frac{1}{6}C_1^{(2)}\right)  && + \quad && D_{d_L}, \notag \\
& m_{\tilde{d}_R}^2    = \quad && m_{16_F}^2 + \quad && m_D^2 \left(-3 + 4C_1^{(1)}\right)
									  && + \quad && m_{1/2}^2 \left(C_3^{(2)} + \frac{2}{3}C_1^{(2)}\right)  && + \quad && D_{d_R}, \notag \\
& m_{\tilde{e}_L}^2    = \quad && m_{16_F}^2  + \quad && m_D^2 \left(-3 - 6C_1^{(1)}\right)
									  && + \quad && m_{1/2}^2\left(C_2^{(2)} + \frac{3}{2}C_1^{(2)}\right)  && + \quad && D_{e_L}, \notag \\
& m_{\tilde{e}_R}^2    = \quad && m_{16_F}^2  + \quad && m_D^2 \left(1 + 12C_1^{(1)}\right)
									  && + \quad &&m_{1/2}^2 \left(6 C_1^{(2)}\right) && + \quad && D_{e_R}, \notag \\
& m_{\tilde{\nu}_L}^2  = \quad && m_{16_F}^2 + \quad && m_D^2 \left(-3 - 6C_1^{(1)}\right)
									  && + \quad && m_{1/2}^2 \left(C_2^{(2)} + \frac{3}{2}C_1^{(2)}\right)  && + \quad &&  D_{\nu_L},
\label{ScalarMassesRGEs}
\end{alignat}
where the $C_a^{(n)}$ are constants, defined as
\begin{equation}
	C_a^{(n)} = 
	\frac{c_a}{b_a} \left( 1 - 
	\frac{g_a^{2n}(M_\text{SUSY})}{g_a^{2n}(M_\text{GUT})} \right),  
	\quad (c_1,c_2,c_3) = \left(\frac{1}{5}, \frac{3}{2},\frac{8}{3}\right),
	\quad (b_1,b_2,b_3) = \left(\frac{33}{5}, 1,-3\right),
\end{equation}
The electroweak D-terms $D_i$ are defined in \eqref{EWDterm} and they are usually sub-dominant to the soft scalar masses. 

\begin{figure}[t]
\centering
\includegraphics[clip,width=0.75\textwidth]{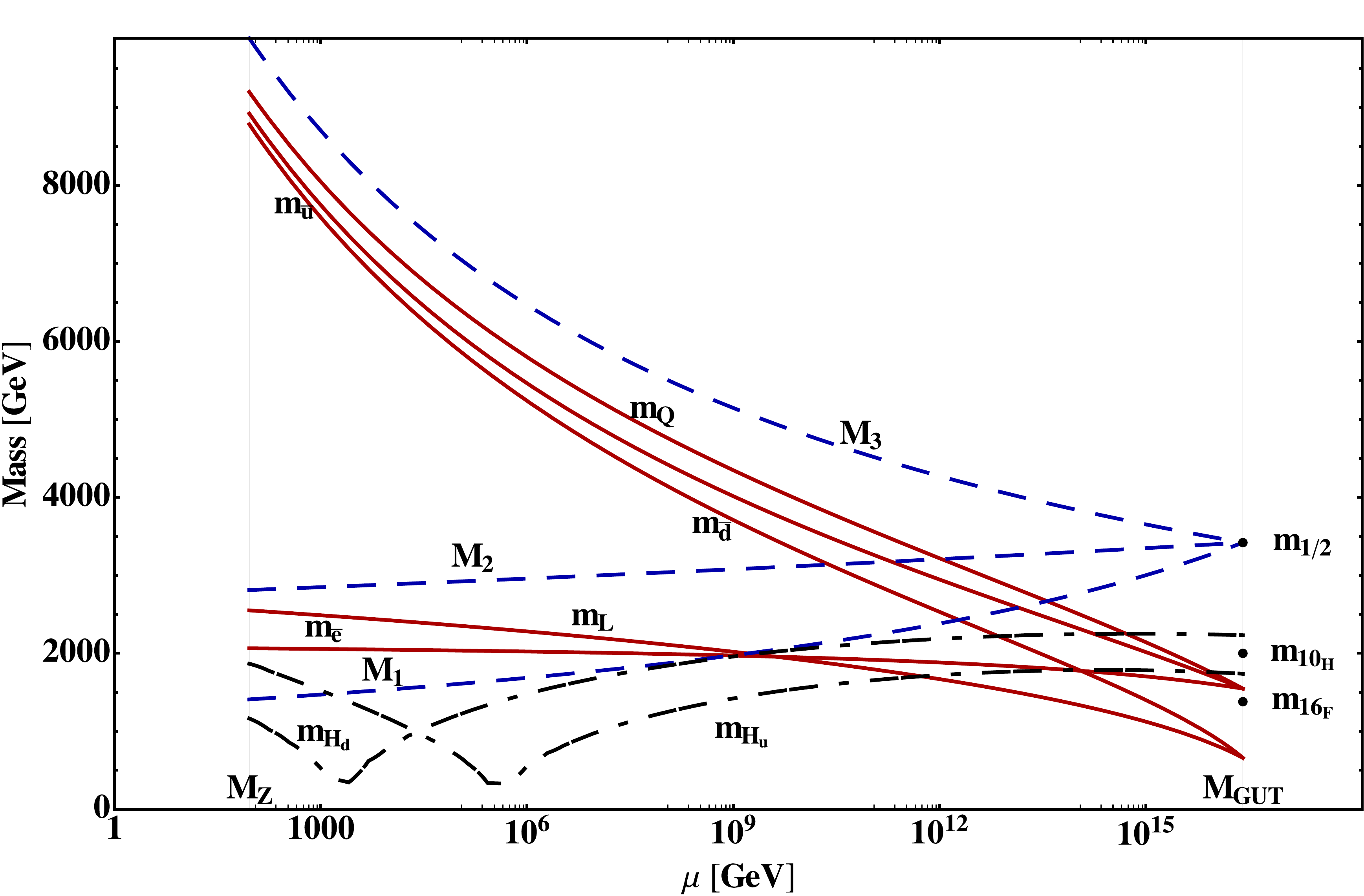}
\caption{Solution of the RGEs for the scalar masses of the 1st generation, the gaugino masses and the Higgs doublet masses in the benchmark scenario defined in equation \eqref{BenchmarkScenario} but with $m_D^2 = (0.7 \TeV)^2$ and $m_{10_H}^2 = (2 \TeV)^2$.}
\label{fig:1genRGEs} 
\end{figure}

The constants $C_a^{(n)}$ depend only on the gauge couplings. However, there is a non-trivial dependence on $\tan\beta$ within the electroweak D-terms. Since they are essentially negligible, we fix $\tan\beta$ to the value in the benchmark scenario described in \eqref{BenchmarkScenario}, $\tan\beta = 39$. The scalar masses for the 1st and 2nd generation squarks and sleptons can then be numerically written as
\begin{align}
	m_{\tilde{u}_L}^2   &=  m_{16_F}^2 + 1.0 m_D^2 + 5.3 m_{1/2}^2 -  (53.6\GeV)^2, \notag \\
	m_{\tilde{u}_R}^2   &=  m_{16_F}^2 + 0.9 m_D^2 + 4.9 m_{1/2}^2 -  (35.8\GeV)^2, \notag \\
	m_{\tilde{d}_L}^2   &=  m_{16_F}^2 + 1.0 m_D^2 + 5.3 m_{1/2}^2 +  (59.3\GeV)^2, \notag \\
	m_{\tilde{d}_R}^2   &=  m_{16_F}^2 - 2.9 m_D^2 + 4.9 m_{1/2}^2 +   (25.3\GeV)^2, \notag \\
	m_{\tilde{e}_L}^2   &=  m_{16_F}^2 - 3.1 m_D^2 + 0.5 m_{1/2}^2 +  (47.3\GeV)^2, \notag \\
	m_{\tilde{e}_R}^2   &=  m_{16_F}^2 + 1.2 m_D^2 + 0.2 m_{1/2}^2 +  (43.9\GeV)^2, \notag \\
	m_{\tilde{\nu}_L}^2 &=  m_{16_F}^2 - 3.1 m_D^2 + 0.5 m_{1/2}^2 -  (64.5\GeV)^2.
\label{MassSquarksandSleptons}
\end{align}
For illustration, Figure \ref{fig:1genRGEs} shows the running of the scalar masses in a representative example scenario. As the usual MSSM RGE running
is driven by the gaugino mass $m_{1/2}$, the additional impact of the SO(10) D-term is roughly determined by the ratio $m_D^2 / m_{1/2}^2$. For $m_D^2 / m_{1/2}^2 \ll 1$, the spectrum will be of the usual CMSSM type, whereas for $m_D^2 / m_{1/2}^2 \gtrsim 1$, the impact of the SO(10) D-term on the sparticle spectrum will be sizeable.

Different sparticle masses in the equations (\ref{ScalarMassesRGEs}, \ref{MassSquarksandSleptons}) depend on the model parameters $m_{16_F}^2$, $m_D^2$ and $m_{1/2}$ with the same or very similar coefficients. We use this to construct linear combinations of these masses that depend on a reduced number of parameters, which will become very useful when trying to find an optimal scenario in the parameter space. The first combination to consider is among the particles belonging to different multiplets in the $\mathrm{SU}(5)$ subgroup of $\mathrm{SO}(10)$. Due to the presence of the D-terms this combination will induce a large splitting between the left and right handed squarks and sleptons, given by
\begin{align}
\label{eq:SquarkSquarkSleptonSleptonSplitting}
	m_{\tilde{d}_L}^2 - m_{\tilde{d}_R}^2  &= \phantom{-}3.9 m_D^2 + 0.4 m_{1/2}^2 + \mathcal{O}(M_Z^2) \notag \\
	m_{\tilde{e}_L}^2 - m_{\tilde{e}_R}^2  &=- 4.3 m_D^2 + 0.3 m_{1/2}^2 + \mathcal{O}(M_Z^2).
\end{align}
Secondly, the splitting between those masses with similar D-term contributions, i.e. those supersymmetric particles that belong to the same multiplet in the $\mathrm{SU}(5)$ subgroup of $\mathrm{SO}(10)$ is given by
\begin{alignat}{3}
\label{SquarkSleptonSplitting}
	m_{\tilde{d}_R}^2 - m_{\tilde{e}_L}^2 &=&  
	&&0.2 m_D^2 + 4.4 m_{1/2}^2 + \mathcal{O}(M_Z^2), \notag \\
	m_{\tilde{u}_L}^2 - m_{\tilde{e}_R}^2 &=& 
	- &&0.2 m_D^2 + 5.1 m_{1/2}^2 + \mathcal{O}(M_Z^2), \notag \\
	m_{\tilde{u}_R}^2 - m_{\tilde{e}_R}^2 &=& 
	- &&0.3 m_D^2 + 4.7 m_{1/2}^2 + \mathcal{O}(M_Z^2).
\end{alignat}
These splittings are largely driven by the gauge contributions proportional to $m_{1/2}$ also present in the CMSSM. Nevertheless, a large SO(10) D-term $m_D^2$ can appreciably contribute to the splitting for small $m_{1/2}$.  

Thirdly, a small splitting is caused by the EW D-terms in the left-handed squarks and the left-handed sleptons, which, belonging to the same $SU(2)$ multiplet, are quasi-degenerate, with a splitting proportional to $M_Z^2$,
\begin{align}
	m_{\tilde{d}_L}^2 - m_{\tilde{u}_L}^2   
	= \mathcal{O}(M_Z^2), \notag \\
	m_{\tilde{e}_L}^2 - m_{\tilde{\nu}_L}^2 = 
	\mathcal{O}(M_Z^2).
\end{align}
The above relations are obtained by using only the 1-loop solution of the RGEs which may not be accurate for large values of $m_D^2$.  We calculate the 2-loop corrections using the approximation discussed in Appendix \ref{app:RGEs} and find that these contributions are, at most,
\begin{align}
  (\delta m^2_\text{2-loop})_{1,2} &< 
  \mathcal{O}(10^{-2}) (- m_{16_F}^2 - m_{1/2}^2) 
  + \mathcal{O}(10^{-3})(- m_{10_H}^2 - m_D^2) , \notag \\
  (\delta m^2_\text{2-loop})_{3} &< 
  \mathcal{O}(10^{-2}) (- m_{16_F}^2 - m_{1/2}^2) 
  + \mathcal{O}(10^{-3})(- m_{10_H}^2 - m_D^2 + A_0^2 + A_0 m_{1/2}),
\end{align}
for the first two and the third generations, respectively. As expected, for large values of the parameters these contributions can be significant and hence we will take them into account in our analysis.

\section{Direct SUSY Searches at the LHC}
\label{sec:direct_susy_searches}

\subsection{Reinterpretation of Squark and Gluino Limits}
\label{sec:reinterpretation}

The most stringent limits on superpartner masses currently come from searches for strongly charged superpartners viz.\ squarks and gluons. LHC searches based on multiple jets and  missing energy currently rule out squarks masses of the order of 2 TeV and gluino masses of the order of 1 TeV depending on the model used for interpretation~\cite{ATLAS-CONF-2013-047, Chatrchyan:2014lfa}.  In this section, we determine how these limits translate to the SUSY SO(10) parameters.

The supersymmetric SO(10) model has two parameters that affect the squark masses at tree level, $m_{16}$ and $m_D^2$.  In particular, a non-zero $m_D^2$ results in a split between left- and right-handed squarks.  Therefore, the simplification in the CMSSM that all squarks of the first two generations are nearly degenerate is lost.  For this analysis, we have retained the universal gaugino sector, meaning the gaugino masses originate from a common parameter at the GUT scale leading to a ratio $M_1:M_2:M_3 = 1:2:6$ at the electroweak scale.

We factorize the problem of estimating final cross section after the cuts into two steps.  Firstly, we analytically calculate the production cross section and the branching fractions.  Secondly, we estimate the efficiencies of the cuts in each production mode for the jets+MET search channels reported by ATLAS using Monte Carlo simulation.

\begin{figure}[tb]
\begin{center}
\includegraphics[width=0.49\textwidth]{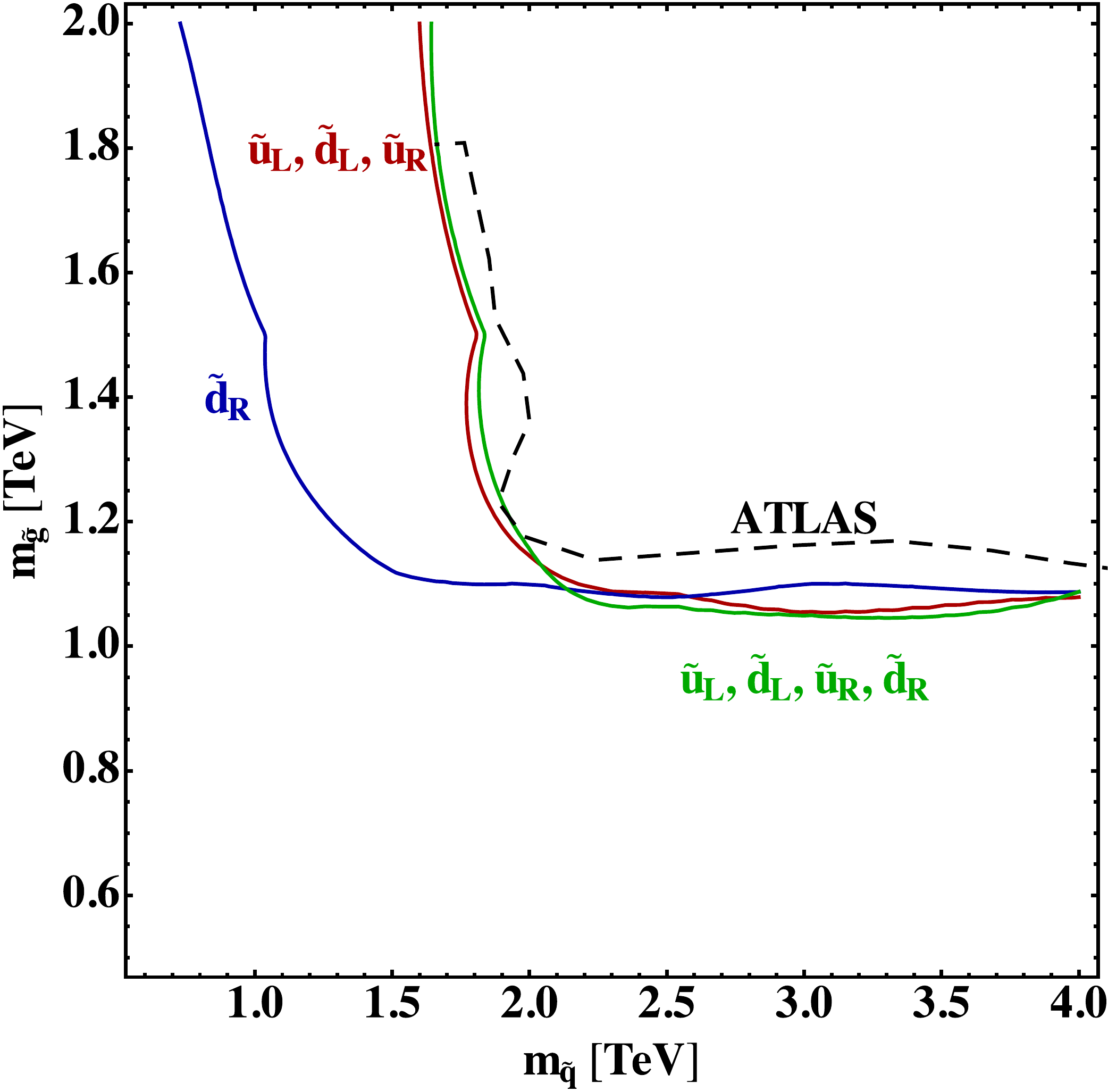}
\end{center}
\caption{Comparison of exclusion limits for CMSSM (green), $m_D^2>0$ (blue) and $m_D^2<0$ (red) simplified models with the ATLAS limit (dashed black). \label{fig:limits}}
\end{figure}
The efficiency of the cuts is calculated using a simplified model with two parameters $m_{\glue}$ and $m_{\sq}$. There are four production modes that result in jets+MET final states viz.\ $\glue \glue$, $\sq \sq$, $\sq \sq^*$ and $\sq \glue$.  We assume each squark decays as $\tilde q \rightarrow q \tilde \chi^0_1$ and the gluino decays via either $\tilde g \rightarrow q \tilde q$ if $m_{\glue} > m_{\sq}$ or via $\tilde g \rightarrow q \bar q \tilde \chi^0_1$ otherwise.  As a consistency check, we reproduce the ATLAS limits based on \cite{ATLAS-CONF-2013-047} for a simplified model where all squarks are degenerate and the lightest (bino-dominated) neutralino is the LSP with a mass a sixth of the gluino mass. The comparison is shown in Figure~\ref{fig:limits}, where the CMSSM model with all squarks being degenerate ($\tilde{u}_L$, $\tilde{d}_L$, $\tilde{u}_R$, $\tilde{d}_R$) is plotted in green and the observed ATLAS limit in dashed black.  The Monte-Carlo simulation was performed using Pythia~8\cite{Sjostrand:2006za,Sjostrand:2007gs, Desai:2011su} with Gaussian smearing of the momenta of the jets and leptons as a theorist's detector simulation. Figure~\ref{fig:limits} demonstrates that we approximately reproduce the exclusion limit reported by ATLAS in our simulation.

To investigate the change in the ATLAS limits given a non-zero $m_D^2$, we use two separate simplified models.  First, corresponding to $m_D^2 \gg 0$, we have the case where right-handed, down-type squarks are much lighter than the rest.  We approximate this by setting $m_{\tilde d_R} = m_{\tilde s_R} = m_{\tilde b_1} = m_{\sq}$ and all other squark masses set to 10 TeV.  Second, corresponding to $m_D^2 \ll 0$, we have the case where all left-handed squarks along with the right-handed up-type quarks are light.  This is approximated by a simplified model where $m_{\sq}$ corresponds to the degenerate mass of all squarks except the ones in the $m_D^2 \gg 0$ model.  The change in the exclusion limit for both of these cases is also shown in Fig.~\ref{fig:limits}, where the $m_D^2 \gg 0$ ($\tilde{d}_R$ light) case is plotted in blue,  and $m_D^2 \ll 0$ case ($\tilde{u}_L$, $\tilde{d}_L$ and $\tilde{u}_R$ light) is in red. The exclusion limit in the case $m_D^2 \ll 0$ is almost identical to the fully degenerate CMSSM case, whereas $m_D^2 \gg 0$ leads to a considerably weaker limit $m_{\tilde q} \gtrsim 1 \TeV$. The gluino limit remains unaffected.

Assuming a similar sensitivity with 300 fb$^{-1}$ integrated luminosity at the 14 TeV run of the LHC, we expect to rule out up to $m_{\sq} \sim 3.2$~TeV for the $m_D^2 \ll 0$ case and $m_{\sq} \sim 2.8$~TeV for the $m_D^2 \gg 0$ case.  The reach in gluino mass is about $m_{\glue} \sim 3.6$~TeV.  A 3-sigma discovery can be made for $m_{\sq} \sim 2.5$~TeV for the $m_D^2 \ll 0$ case and $m_{\sq} \sim 1.8$~TeV for the $m_D^2 \gg 0$ case. We have added a comment
in this regard in section 2.3.

\subsection{Summary of other LHC SUSY Searches}
\label{sec:LHC_searches}

After the first run of the LHC, a great amount of the data has been analyzed and comprehensive searches for supersymmetric signals have been carried out. Both ATLAS and CMS have done an extensive survey of many different scenarios and studied the data collected in the most model independent way possible, so as to exclude as much of the SUSY parameter space as possible. We summarize here the exclusion limits for some of the supersymmetric particles:

\paragraph{Stops and Sbottoms} 
Stops are produced at the LHC mostly through the s-channel, and the primary decay modes are $\st \to t \neut$ and $\st \to b \chargpm$. The final states studied have the signature $4j + l + MET$, with none to three $b-$tags and the current lower limit on the stop mass is around $m_{\tilde t} \gtrsim 650 \GeV$. However, if the stop is not allowed to decay to an on-shell top, $m_{\st} < m_t + m_{\neut}$, the decay phase space is reduced and the process is suppressed which weakens the limit to $m_{\tilde t} \gtrsim 250 \GeV$. Searches for sbottoms are similar to those for stops, with similar production rates and complementary decays, $\sb \to b\neut$ and $\sb \to t \chargpm$. Consequently, the mass limits are similar, $m_{\tilde b} \gtrsim650 \GeV$ \cite{ATLAS-CONF-2013-024, Aad:2013ija, Chatrchyan:2013xna, CMS-PAS-SUS-13-008, Chatrchyan:2013lya}.

\paragraph{Sleptons, Neutralinos and Charginos}
Although electroweak processes at the LHC are several orders of magnitude smaller than strong ones, the precision of the measurements done by ATLAS and CMS is good enough to provide a limit of $m_{\tilde l} \gtrsim 300 \GeV$. Similar to the sleptons, the limits on the neutralinos and charginos are considerably weaker than those of gluinos and squarks. Using purely electroweak processes such as $\tilde \chi_2^0 \chargpm \to Z \neut W^\pm\neut$ or $\tilde \chi^0_2 \chargpm \to l \tilde{\nu} \tilde{l} l (\nu \tilde{\nu})$, both LHC experiments have currently excluded masses up to $m_{\tilde\chi} \gtrsim 300 \GeV$ \cite{ATLAS-CONF-2013-028, ATLAS-CONF-2013-049, CMS-PAS-SUS-13-006, CMS-PAS-SUS-13-017}. Finally, the extra Higgs states predicted by supersymmetry have also been subject to scrutiny.  However, due to the strong dependence on the parameters in the MSSM (particularly $\tan\beta$), the limits are not very strong. As of today, the limits seem to favour $\tan\beta \gtrsim 18$ and Higgs masses around or above that of the found Higgs state, $m_{H,A,H^\pm} \gtrsim 100 \GeV$ \cite{Aad:2012cfr, ATLAS-CONF-2013-090, Chatrchyan:2012vca,CMS-PAS-HIG-13-021}.

\section{Analysis}
\label{sec:Analysis}

The SUSY $\mathrm{SO}(10)$ model has 7 free parameters, $m_{16_F}^2, m_{10_H}^2, m_{1/2}, m_D^2, A_0, \tan \beta, \text{sign}(\mu)$, when no constraints are imposed.  We will use existing experimental limits to fix or constrain some of these model parameters using the results of section \ref{sec:direct_susy_searches}, focusing on the most interesting deviations from the standard CMSSM scenario.

As discussed above, there is a lower limit on the mass of the lightest squark, at $m_{\tilde{q}} \gtrsim 2 \TeV$ within the framework of the CMSSM. With the degeneracy of all scalar particles at the GUT scale, this bound also forces the sleptons to become heavy, usually well beyond the direct detection slepton mass limits. However, in the minimal SUSY SO(10) model, it is possible to evade the squark limits while keeping the slepton masses light, possibly at the level of experimental detectability. We will therefore seek to explore the model parameter space with a large splitting between the squark and slepton masses by taking advantage of the relation \eqref{SquarkSleptonSplitting}. Even in the CMSSM, one may obtain relatively light sleptons (compared to squarks) by increasing the RG running effect of the strong gauge coupling by increasing $m_{1/2}$. A large value of $m_{1/2}$ is actually required due to the corresponding gluino mass limit $m_{\tilde g} \gtrsim 1\TeV$. For a fixed squark mass, this approach has the disadvantage that it will also raise the  lightest neutralino mass which is the preferred Lightest Supersymmetric Particle (LSP) candidate. In order to have the lightest neutralino lighter than any charged sparticle for as much of the parameter space as possible, we will fix the value of $m_{1/2}$ so as to produce a gluino with a mass roughly at the current limit, $m_{\tilde{g}} \approx 1 \TeV$. 

The only other free parameter in \eqref{SquarkSleptonSplitting} is $m_D^2$, which has a comparatively small contribution towards the splitting. This is because the the scalar species under consideration belong to the same SU(5) multiplets and the splitting is caused by a secondary effect in the RGEs. Notice also that the splitting for the $\mathbf{\bar 5}$ and $\mathbf{10}$ multiplets has opposite signs in their dependence on $m_D^2$, cf. \eqref{eq:SquarkSquarkSleptonSleptonSplitting}, i.e. for $m_D^2 \gg 0$, $\tilde{e}_L$, $\tilde{d}_R$ will be the lighter states whereas for $m_D^2 \ll 0$ it will be $\tilde{e}_R$ and $\tilde{u}_L$.

We will therefore look for a region of parameter space where, by increasing $m_D^2$ in both positive and negative directions, we achieve a large splitting between squarks and sleptons. Since $m_{1/2}$ is fixed, as stated above, and in order to keep the mass of the lightest first generation squark ($m_{\tilde{q}}$) fixed to the lowest allowed value, we express $m_{16_F}^2$ as a function of the other model parameters and the desired squark mass $m_{\tilde{q}}$,
\begin{equation}
	m_{16_F}^2 = 
	m_{\tilde{q}}^2 - c_1 m_D^2 - c_2 m_{1/2}^2 - c_3 + \delta_2,
\end{equation}
where the constants $c_i$ are taken from \eqref{MassSquarksandSleptons} for the corresponding squark species and $\delta_2$ is the 2-loop correction to the mass of the lightest squark. The latter is significant for large $|m_D^2|$ and $m_{16_F}^2$. The limit of this procedure is reached as soon as one of the particles becomes tachyonic (negative squared mass) at the electroweak scale.

Due to the large third generation Yukawa couplings, especially for the top quark, the third generations of sparticles are usually lighter than the first two. We will consider this case first in the following section. In section \ref{sec:Light1gen}, we will describe the possibility of having the first two generations lighter than the third by compensating the RG effect of the Yukawa couplings. To conclude, in section \ref{sec:NonUniversalGauginosAnalysis}, we will study the additional impact of non-universal gauginos on the sparticle spectrum.

\subsection{Light Third Generation}
\label{sec:Light3gen}

%
\begin{figure}[t]
\centering
\includegraphics[width=0.75\textwidth]{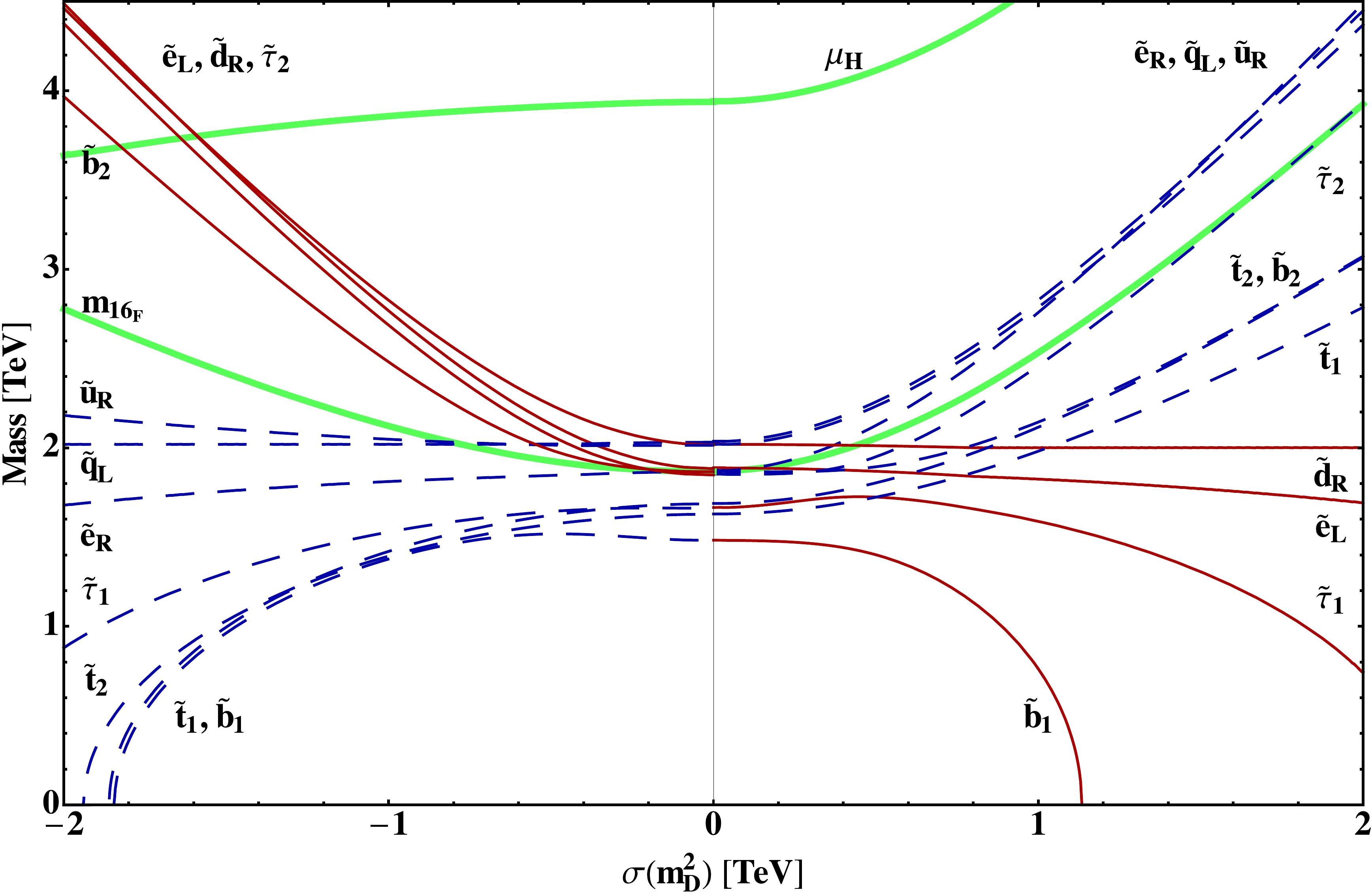}
\caption{Sparticle masses as a function of $\sigma(m_D^2) = \sign(m_D^2) \sqrt{|m_D^2|}$. The remaining model parameters are fixed as described in Eq.~\eqref{eq:BenchmarkLightThird}.}
\label{fig:Light3gen}
\end{figure}
Starting with the benchmark scenario described in \eqref{BenchmarkScenario}, and parameters set by the current LHC limits we will perform a scan over $m_D^2$ to analyse how the masses of different sparticles behave. To achieve a light but viable SUSY spectrum, the value of $m_{1/2}$ is fixed such that $m_{\tilde{g}} = 1 \TeV$ at the current exclusion limit. The value of $m_{16_F}^2$ is then determined so as to keep the lightest squark at a mass of $2$ TeV for a given $m_D^2$. Please note that while the limit on the squark mass is reduced for $m_D^2 \gg 0$, cf. section \ref{sec:reinterpretation}, we will use $m_{\tilde q} = 2 \TeV$ in all cases for easy comparison. The remaining model parameters are thus fixed as
\begin{gather}
  m_{10_H}^2   = -(3647 \GeV)^2, \quad
  m_{1/2}      = 389    \GeV,    \notag\\ 
  A_0          = -3140  \GeV,    \quad
  \tan\beta    = 39,             \quad
  \sign(\mu_H) = 1,              \notag\\
  m_{16_F}^2 \text{ such that } \min(m_{\tilde q}) = 2\TeV,
\label{eq:BenchmarkLightThird}
\end{gather}
unless otherwise noted. Figure \ref{fig:Light3gen} shows the dependence of the masses on $m_D^2$ for both scenarios, using the 2-loop RGEs described in Appendix \ref{app:RGEs}. Most obviously, the splitting between the sparticles in different representations of SU(5) increases with larger values of $|m_D^2|$. However the splitting between the first generation squarks and sleptons does not get big enough for the sleptons to become appreciably lighter before the third generation stops and sbottoms become tachyonic. For both $m_D^2 > 0$ and $m_D^2 < 0$, the lightest sparticle is the lightest sbottom. The regions with $m_D^2 \gtrsim (1.1 \TeV)^2$ and $m_D^2  \lesssim -(1.8\TeV)^2$ are non physical. For the case of negative $m_D^2$ we have obtained, in a rather natural way, very light stops, sbottoms and staus, while the rest of the scalars are above $1 \TeV$. This is consistent with current experimental data~\cite{ATLAS-CONF-2013-024, Chatrchyan:2013xna} and would provide a natural solution to the hierarchy problem, with a reasonable fine tuning due to light stops and sbottoms. We have, however, chosen a mass for the gluino fixed at 1 TeV resulting in relatively light neutralinos, $m_{\tilde{\chi}^0_1} \approx 150 \GeV$. In addition to the low energy sparticle masses, Figure \ref{fig:Light3gen} also shows the derived value of the Higgs $\mu_H$ term, and the soft mass $m_{16_F}$ at the GUT scale, respectively. An example sparticle spectrum for this scenario is shown in Figure~\ref{fig:Spectrum}~(left) for $m_D^2 = -(1.83 \TeV)^2$.

\begin{figure}[t]
\centering
\includegraphics[width=0.49\textwidth]{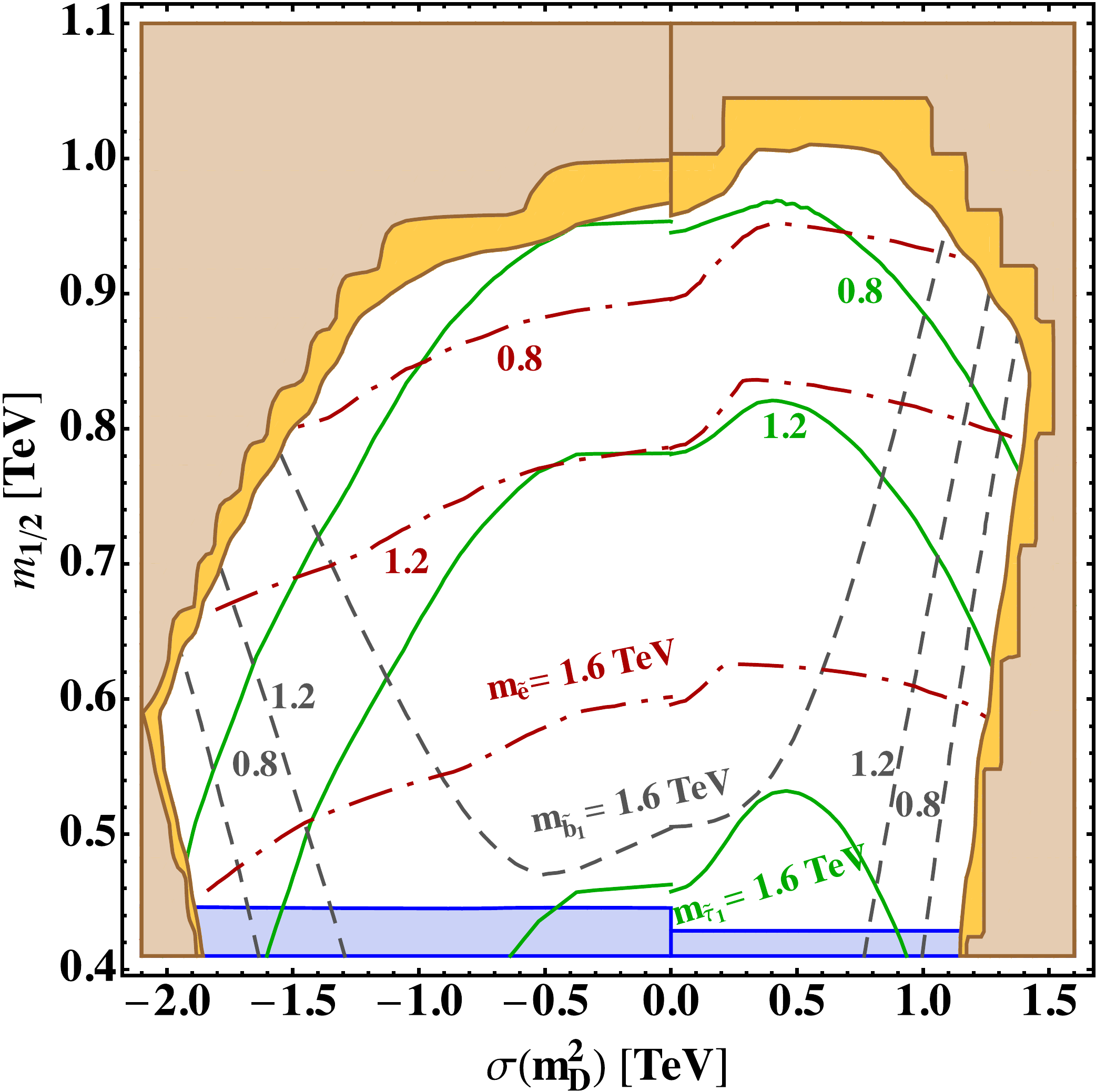}
\includegraphics[width=0.49\textwidth]{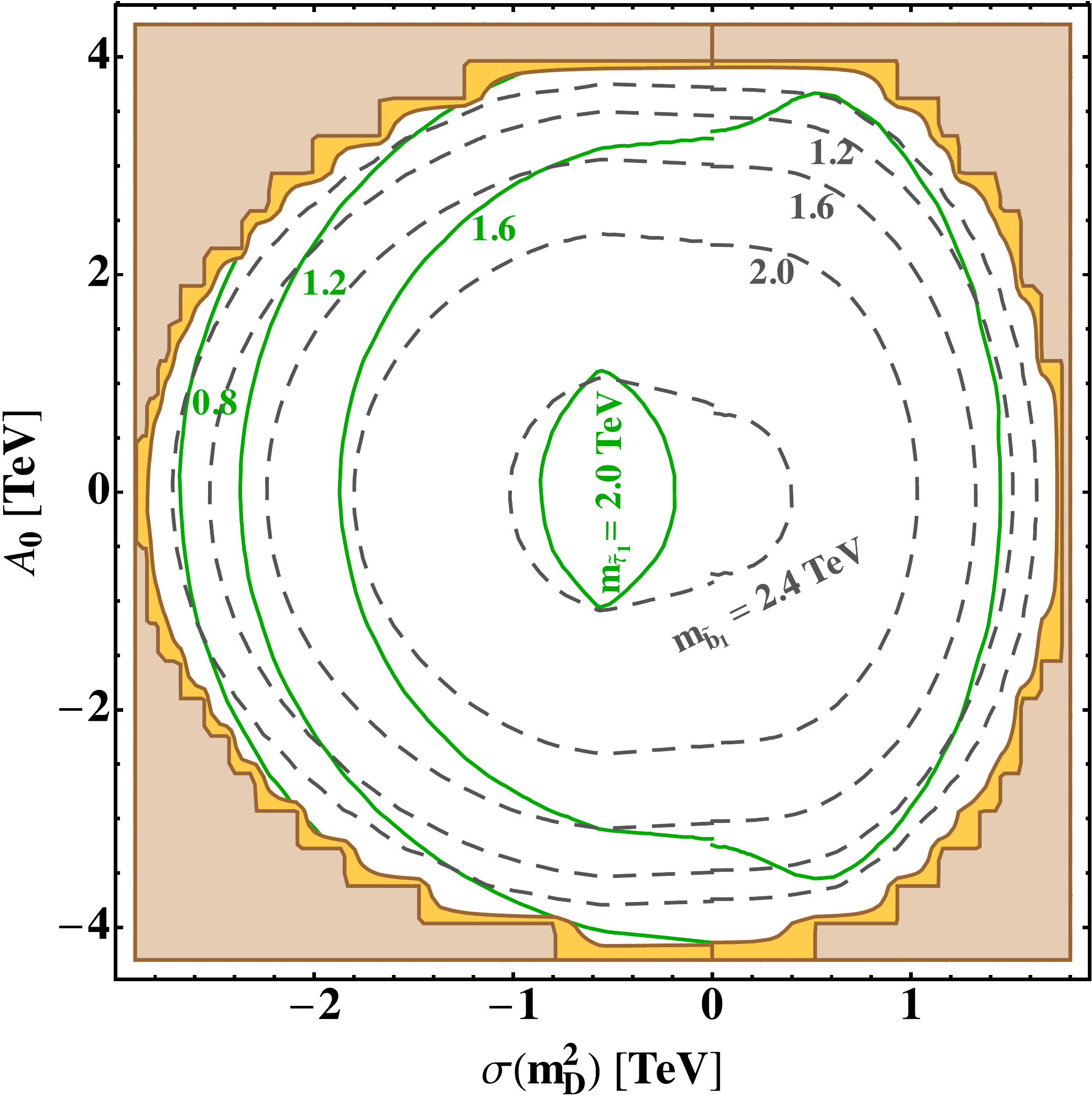}
\caption{Mass of the lightest stau $\tilde{\tau}_1$ (solid green), sbottom $\tilde{b}_1$ (dashed grey) and selectron $\tilde{e}$ (dash-dotted red) as a function of of $m_D^2$ and $m_{1/2}$ (left) and of $m_D^2$ and $A_0$ (right). The remaining parameters are respectively fixed as described in Eq.~\eqref{eq:BenchmarkLightThird}. The coloured areas are excluded or disfavoured because there is at least one tachyonic state (brown), the neutralino is not the LSP (orange), the gluino mass is below the experimental limit (blue).}
\label{fig:Light3gen-mDm12-mDA0}
\end{figure}
The impact of different values for $m_{1/2}$ can be seen in Figure~\ref{fig:Light3gen-mDm12-mDA0}~(left) where the allowed $(m_D^2, m_{1/2})$ space is shown. Also displayed are the lightest slepton $m_{\tilde{e}}$, the lightest stau $m_{\tilde{\tau}_1}$ and the lightest sbottom $m_{\tilde{b}_1}$ mass. The outer, shaded (brown) area is excluded because there is at least one tachyonic state, usually the sbottom. The enclosing (orange) band denotes the parameter space where the neutralino $\tilde{\chi}_1^0$ is not the LSP. The bottom (blue) band is excluded by the  gluino mass limit from the direct searches described in section \ref{sec:direct_susy_searches}, ($m_{\tilde{g}} \gtrsim 1.1\TeV$). We can clearly see that increasing $m_{1/2}$ has the effect of lowering the masses of all the affected sparticles, particularly the sleptons, cf. Eq.~\eqref{SquarkSleptonSplitting}. However, the mass of the lightest neutralino increases with $m_{1/2}$, and for $m_{\tilde\chi^0_1} \approx 0.4\TeV$, one of $\tilde\tau_1$, $\tilde e$ or $\tilde b_1$ becomes lighter. For $m_{1/2}$ close to the upper limit, $m_{1/2} \approx 0.9 \TeV$, either the the lightest stau or selectron is the NLSP.

In order to have a better understanding why the third generation squarks are so light compared to their first and second generation counterparts, Figure~\ref{fig:Light3gen-mDm12-mDA0}~(right) displays the corresponding properties in the $(m_D^2,A_0)$ parameter plane. Notice that for the sbottom and the stau, the effects of large $m_D^2$ and large $A_0$ are similar, i.e. they both push the masses down. As a matter of fact, we can actually see that the sbottom is only the lightest for large $A_0$ (as was the case in Figure \ref{fig:Light3gen}), but is heavier than the stau for small $A_0$, and can even be rather heavy ($m_{\tilde b_1} \approx 2.4 \TeV$). The effect of $A_0$ on the first and second generation slepton mass is negligible due to the small Yukawa couplings, and we do not show it in the plot.

\subsection{Light First Generation}
\label{sec:Light1gen}

\begin{figure}[t]
\centering
\includegraphics[width=0.75\textwidth]{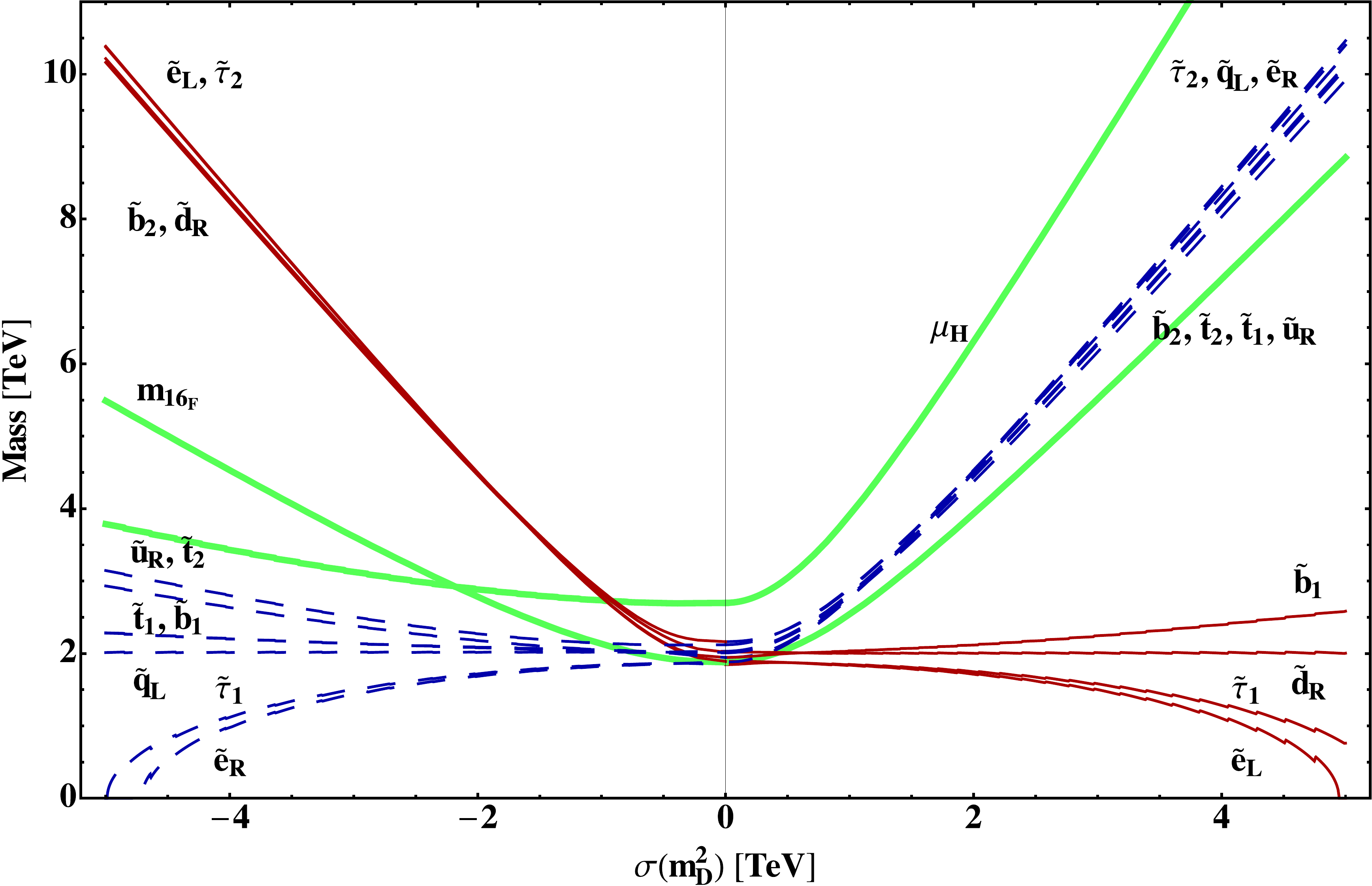}
\caption{As Figure~\ref{fig:Light3gen}, but with the remaining model parameters fixed as described in Eq.~\eqref{eq:BenchmarkLightFirst}.}
\label{fig:Light1gen}
\end{figure}

As described above, the lightest sbottom and stop generically constitute the lightest sferm\-ion states, except for large values of $m_{1/2}$ and $|m_D^2|$. The well known reason for this suppression, also with respect to the first two squark generations, are the large third generation Yukawa couplings which drive the masses down through RGE running. If we look into the terms in the RGEs proportional to the Yukawa couplings (see Appendix \ref{app:RGEs}), we find that they have the following dependence at the one loop level, 
\begin{equation}
	\Delta_{\tau,b,t} \propto m_{10_H}^2 + 2 m_{16_F}^2 + A_0^2.
\label{3genshift}
\end{equation}
Hence, in order to minimize this contribution, we need to compensate the increasingly large values of $m_{16_F}^2$ with equally large and opposite sign values of $m_{10_H}^2 + A_0^2$. If we want to keep the trilinear couplings real, the best choice for this would be $A_0 = 0$ and $m_{10_H} ^2= - 2 m_{16_F}^2$. Including two loop corrections to the masses, one needs to increase this proportionality by about $5 - 10\%$ to compensate the suppression of the stau, stop and sbottoms masses with respect to the first two generations. In the following we will use the relation $m_{10_H}^2 = - 2.1 m_{16_F}^2$. This clearly defines a rather fine-tuned solution as the Yukawa couplings are a priori unrelated to the soft SUSY breaking parameter. We nevertheless study this case as an extreme departure from the generic picture described in section~\ref{sec:Light3gen}. In summary, the base model parameters used in this section are described by
\begin{gather}
  m_{1/2}      = 389    \GeV,    \notag\\ 
  A_0          = 0,    \quad
  \tan\beta    = 39,             \quad
  \sign(\mu_H) = 1,              \notag\\
  m_{16_F}^2, m_{10_H}^2 = -2.1m_{16_F}^2 \text{ such that } 
  \min(m_{\tilde q}) = 2\TeV,
\label{eq:BenchmarkLightFirst}
\end{gather}
unless otherwise noted. Figure~\ref{fig:Light1gen} shows the effect of approximately compensating the third generation Yukawa couplings on the sparticle masses. We see that indeed the third generation sparticles are heavier than their first generation counterparts. In comparison with Figure~\ref{fig:Light3gen}, the SO(10) D-term $m_D^2$ can be larger, up to $m_D^2 \lesssim (5 \TeV)^2$, in turn producing a wider splitting between the lightest squarks and the lightest sleptons. On the other hand, the heavy squarks and sleptons would be split off considerably, with masses  up to $10\TeV$. This is a clear example of a Split-SUSY \cite{ArkaniHamed:2004fb, Giudice:2004tc, ArkaniHamed:2012gw, Arvanitaki:2012ps} scenario, exhibiting a three-fold splitting: Very light sleptons $\approx 0.1 - 0.2 \TeV$, lightest squarks around $2 - 4 \TeV$ and very heavy squarks and sleptons at  $9 - 10 \TeV$. An example sparticle spectrum for this scenario is shown in Figure~\ref{fig:Spectrum}~(right) for $m_D^2 = +(4.87 \TeV)^2$.
\begin{figure}[t]
\centering
\includegraphics[width=0.49\textwidth]{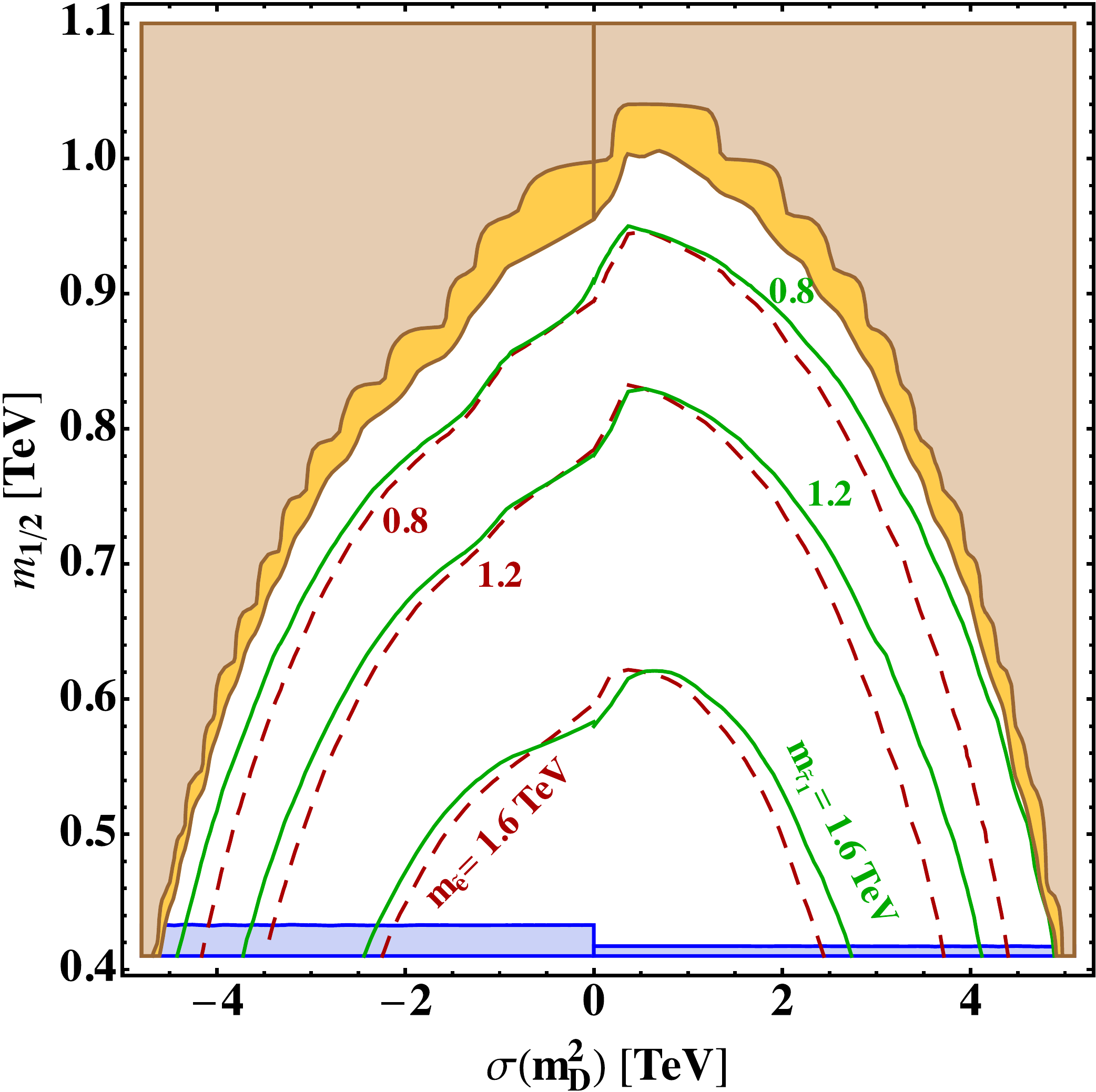}
\includegraphics[width=0.49\textwidth]{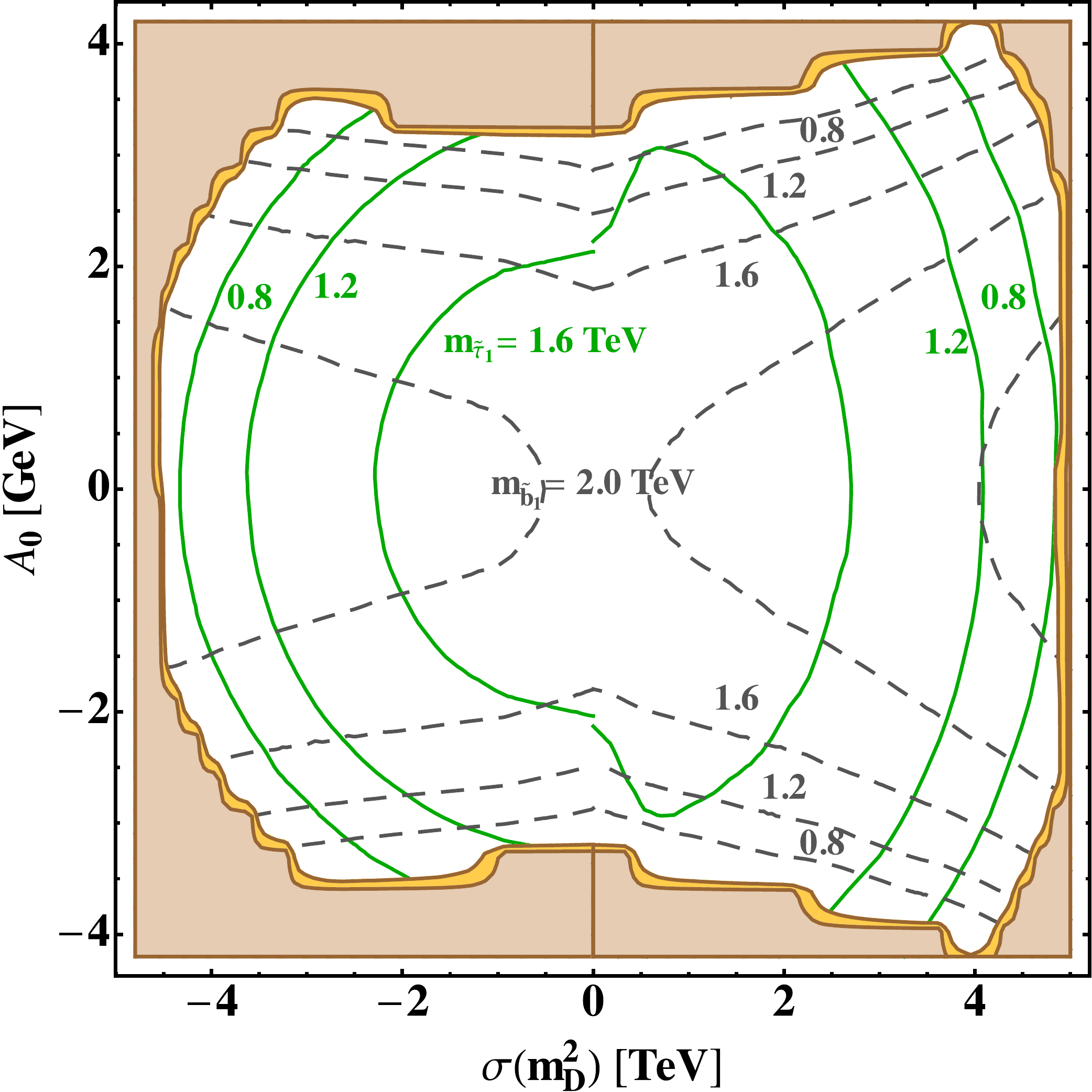}
\caption{As Figure~\ref{fig:Light3gen-mDm12-mDA0}, but with the remaining model parameters fixed as described in Eq.~\eqref{eq:BenchmarkLightFirst}.}
\label{fig:Light1gen-mDm12-mDA0}
\end{figure}
The combined dependencies on $m_D^2$ and either $m_{1/2}$ or $A_0$ is displayed in Figure~\ref{fig:Light1gen-mDm12-mDA0}. The excluded or disfavoured shaded areas are defined as before in Figure~\ref{fig:Light3gen-mDm12-mDA0}. We do not plot the lightest sbottom mass in Figure~\ref{fig:Light1gen-mDm12-mDA0}~(left) as it is too heavy to be of interest here. The main difference from the light third generation case displayed in Figure~\ref{fig:Light1gen-mDm12-mDA0} is that the first generation sleptons are slightly lighter than the light stau, except for small values of $|m_D^2|$. Due to the potentially higher values of $|m_D^2|$, very small slepton masses are possible even for low values of $m_{1/2}$.

The dependence on $A_0$, Figure~\ref{fig:Light1gen-mDm12-mDA0}~(right) in this case is also rather different from Figure~\ref{fig:Light3gen-mDm12-mDA0}~(right). While the stau mass exhibits a similar behaviour, the sbottom mass becomes heavier with increasing $|m_D^2|$ but lighter with increasing $A_0$. This is expected as we do not compensate the effect of $A_0$ on the Yukawa-driven RGE contributions. As a consequence, the lightest sbottom will become the lightest sfermion for large $A_0 \gtrsim 3\TeV$.

The scenario described here would be optimal for sleptons searches at LHC because it allows for very light first, second and also third generation sleptons. Naively, one might expect that the presence of very light (left-handed) smuons is able to enhance the predicted value of the anomalous magnetic moment of the muon closer to the experimentally favoured value, $\Delta a_\mu \equiv a_\mu^\text{exp} - a_\mu^\text{SM} = (26.1 \pm 8.0) \times 10^{-10}$ \cite{Davier:2010nc}. This is because the supersymmetric contributions to $a_\mu$ are driven by muon sneutrino-chargino and smuon-neutralino loops. Unfortunately, the SUSY scenarios considered here require a large Higgs $\mu$-term $\mu_H$ as shown in Figures~\ref{fig:Light3gen} and  \ref{fig:Light1gen}. For a strongly split scenario as in our case, the SUSY contribution is roughly \cite{Moroi:1995yh, Endo:2013bba}
\begin{align}
	\Delta a_\mu^\text{SUSY} \lesssim 
	10^{-8} \times \frac{\tan\beta}{10}
	\frac{(100\GeV)^2}{M_1 \mu_H},
\end{align}
with the lightest gaugino mass $M_1$. Consequently, a strongly split scenario with large $|m_D^2|$ in minimal SUSY SO(10) does not enhance $\Delta a_\mu^\text{SUSY}$ appreciably compared to the standard CMSSM case. 

\subsection{Non-Universal Gauginos}
\label{sec:NonUniversalGauginosAnalysis}

As a final step of our analysis, we will briefly comment on the impact of non-universal gauginos at the GUT scale. In Table \ref{tab:GauginoBC} we see that there are three representative cases: (a) The messenger field is in the singlet representation of the SU(5) embedded in SO(10). This corresponds to the standard universal case with an approximate gaugino hierarchy of $|M_1| : |M_2| : |M_3| = 1/6 : 1/3 : 1$ near the EW scale, which we have discussed above. (b) The messenger is in the $\mathbf{24}$-dimensional representation. Here, the bino is comparatively lighter than in the CMSSM, with an approximate gaugino hierarchy of $|M_1| : |M_2| : |M_3| = 1/12 : 1/2 : 1$ near the EW scale. This is phenomenologically interesting as it creates a larger splitting between the lightest neutralino (essentially the bino) and the gluino. It potentially permits a very light neutralino while satisfying the direct gluino mass limits, cf. section~\ref{sec:direct_susy_searches}. For example, for a gluino mass at the current limit, $m_{\tilde g} \approx 1.1 \TeV$, the lightest neutralino could be lighter than $m_{\tilde\chi_1^0} \approx 100 \GeV$, subject to direct search limits, cf. section~\ref{sec:LHC_searches}. On the other hand, the ratio between $M_2$ and $M_3$ is smaller than that of normal CMSSM, making the second neutralino and lightest chargino slightly heavier. Such a change will for instance suppress the SUSY contribution to the anomalous magnetic moment of the muon. The largest contribution comes from a sneutrino-chargino loop, and the experimental situation would prefer both the SU(2) gaugino and the sleptons to be light. (c) The messenger is in the $\mathbf{200}$-dimensional representation, corresponding to a low energy hierarchy $|M_1| : |M_2| : |M_3| = 5/3 : 2/3 : 1$. The spectrum is rather different here, with the bino being the heaviest gaugino, while the mass of the wino is approximately $2/3$ of the gluino mass. Hence, the lightest neutralino would be mostly wino and would have a relatively large mass for a given gluino mass, compared to the previous case.

\begin{figure}[t]
\centering
\includegraphics[width=0.7\textwidth]{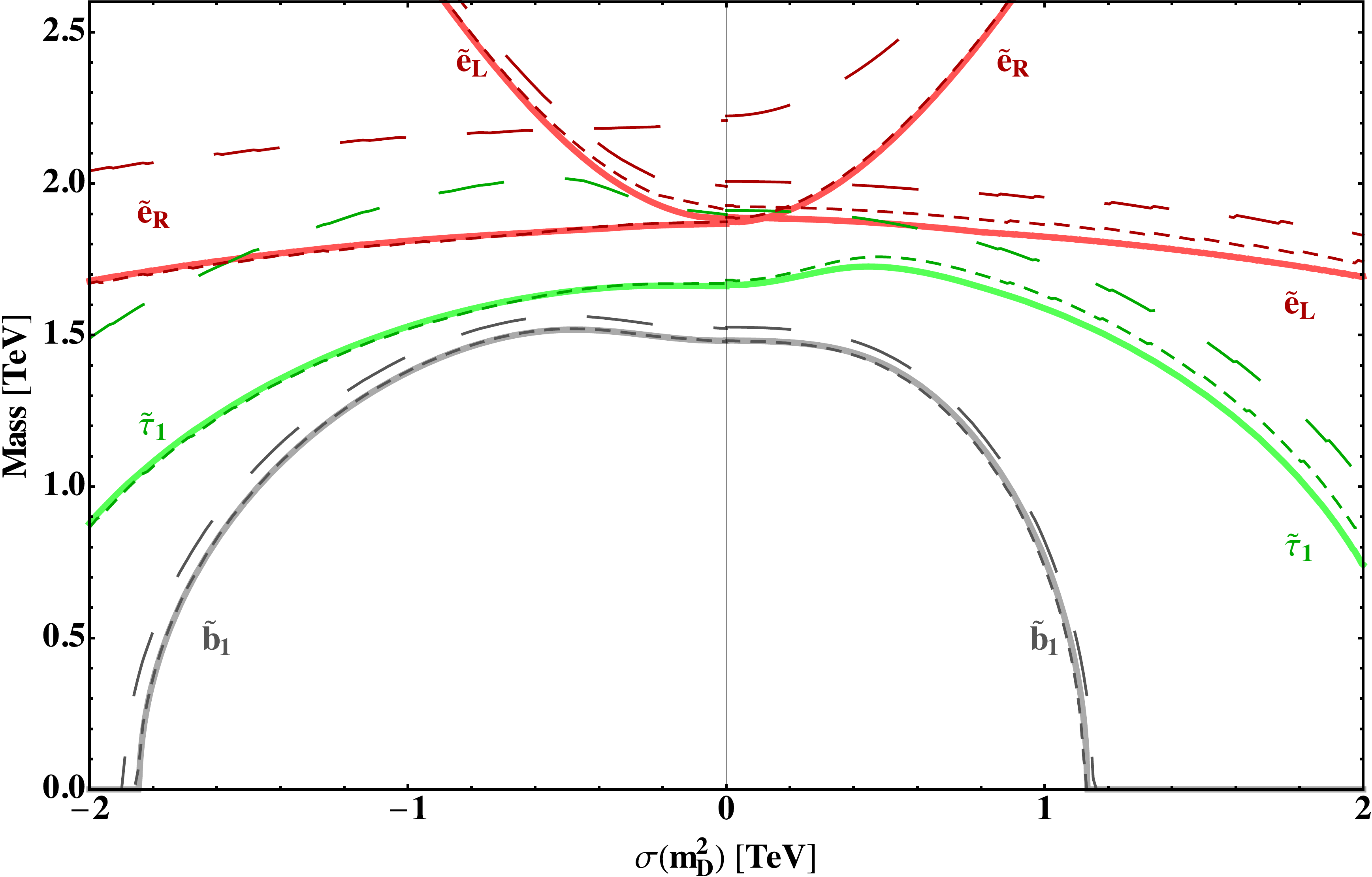}
\caption{Sparticle masses as a function of $\sigma(m_D^2) = \sign(m_D^2) \sqrt{|m_D^2|}$. The remaining model parameters are fixed as described in Eq.~\eqref{eq:BenchmarkLightThird} for three different gaugino hierarchies at the GUT scale: (a) $M_1 = M_2 = M_3 = m_{1/2}$ (universality, solid); (b) $-2 M_1 = -3/2 M_2 = M_3 = m_{1/2}$ (light bino, short dashed); (c) $10 M_1 = 2 M_2 = M_3 = m_{1/2}$ (light wino, long dashed).}
\label{fig:NonUniversalGauginos}
\end{figure}
Other than the direct effect on the gaugino masses, the presence of non-universal gauginos at the GUT scale will also affect the masses of the scalar SUSY particles due to the impact on the RGE running. So far we have calculated scalar particle masses assuming degenerate gauginos at the GUT scale, resulting in a term $\propto m_{1/2}^2$ as the main RGE effect on the scalar masses, see for example Eq.~\eqref{SquarkSleptonSplitting}. Allowing for arbitrary individual gaugino masses $M_1$, $M_2$ and $M_3$ at the GUT scale, these equations will take the form 
\begin{align}
	m_{\tilde{d}_R}^2 - m_{\tilde{e}_L}^2 &=  
	\phantom{-} 0.2 m_D^2 - 0.02 M_1^2 - 0.5 M_2^2 + 4.9 M_3^2 
	+ \mathcal{O}(M_Z^2), \notag \\
	m_{\tilde{u}_L}^2 - m_{\tilde{e}_R}^2 &= 
	-0.2 m_D^2 - 0.15 M_1^2 + 0.5 M_2^2 + 4.9 M_3^2 
	+ \mathcal{O}(M_Z^2), \notag \\
	m_{\tilde{u}_R}^2 - m_{\tilde{e}_R}^2 &= 
	-0.3 m_D^2 - 0.08 M_1^2 \phantom{+ 0.5 M_2^2}\,\,\, + 4.8 M_3^2 
	+ \mathcal{O}(M_Z^2).
	\label{eq:SquarkSleptonSplittingNonUni}
\end{align}
By far the largest contribution is due to the strong gauge effect of the gluino affecting the squarks. In fixing the gluino mass as $m_{\tilde g} \approx 1.1 \TeV$ in tune with the experimental bound, we essentially set the scale of the absolute squark masses. The gaugino non-universality will then induce an additional splitting between the squarks and sleptons, dominantly driven by the wino mass $M_2$.
A comparison of the three cases is shown in Figure~\ref{fig:NonUniversalGauginos}, i.e. (a) universal gauginos (solid),  (b) light bino case (short dashed) and (c) light wino case (long dashed). As expected from \eqref{eq:SquarkSleptonSplittingNonUni}, case (b) produces only small deviations when compared to universal gauginos. On the other hand, case (c) can have a sizable impact on the slepton masses, especially for $m_D^2 < 0$. The negative signs in front of $M_1^2$ in \eqref{eq:SquarkSleptonSplittingNonUni} explain the larger slepton masses compared to the universal gaugino case.

\section{Conclusions}
\label{sec:conclusions}

Supersymmetric models are feeling the pinch from the lack of new physics signals at the LHC and in low energy observables. While any phenomenological limits can be evaded by sending the SUSY particle masses to higher scales, such a solution will usually negate the ability of many SUSY models to solve the hierarchy problem of the Standard Model. Minimal scenarios, such as the CMSSM are especially difficult in this regard as the stringent lower limits from LHC direct searches on coloured states will similarly affect non-coloured sparticles. As a consequence, there is now much effort going into the study of less constrained models of low energy SUSY with a large variety of spectra. For example, phenomenological approaches like the phenomenological MSSM do not contain a priori relations between different sparticle species.

In this work, we focused on the other hand on a minimal supersymmetric SO(10) model incorporating one-step symmetry breaking from SO(10) down to the Standard Model gauge group at the usual GUT scale. Such SUSY GUT scenarios are of course very well motivated with the possibility of unifying the gauge and Yukawa couplings at the GUT scale. With respect to the SUSY spectrum, the GUT unification also provides a motivation for the degeneracy of the soft SUSY breaking masses and couplings. In contrast to the CMSSM though, the scalar masses in an SO(10) GUT are shifted by D-terms associated with the breaking of SO(10) to the lower-rank SM group. These D-terms do depend on the details of the gauge breaking but are generally expected to be of the order of the SUSY breaking scale (for example described by SUSY breaking mass $m_{16_F}^2$ of the matter SO(10) $\mathbf{16}$-plet), and can be parametrized by a single additional quantity $m_D^2$. This provides a controlled departure from the degeneracy of the CMSSM. In addition, we also briefly discuss the possibility of non-universal gaugino masses at the GUT. This is a general possibility in SUSY GUT models with gravity mediated breaking if the SUSY breaking messenger is not a singlet under the GUT gauge group. 

\begin{figure}[t]
\centering
\includegraphics[width=0.49\textwidth]{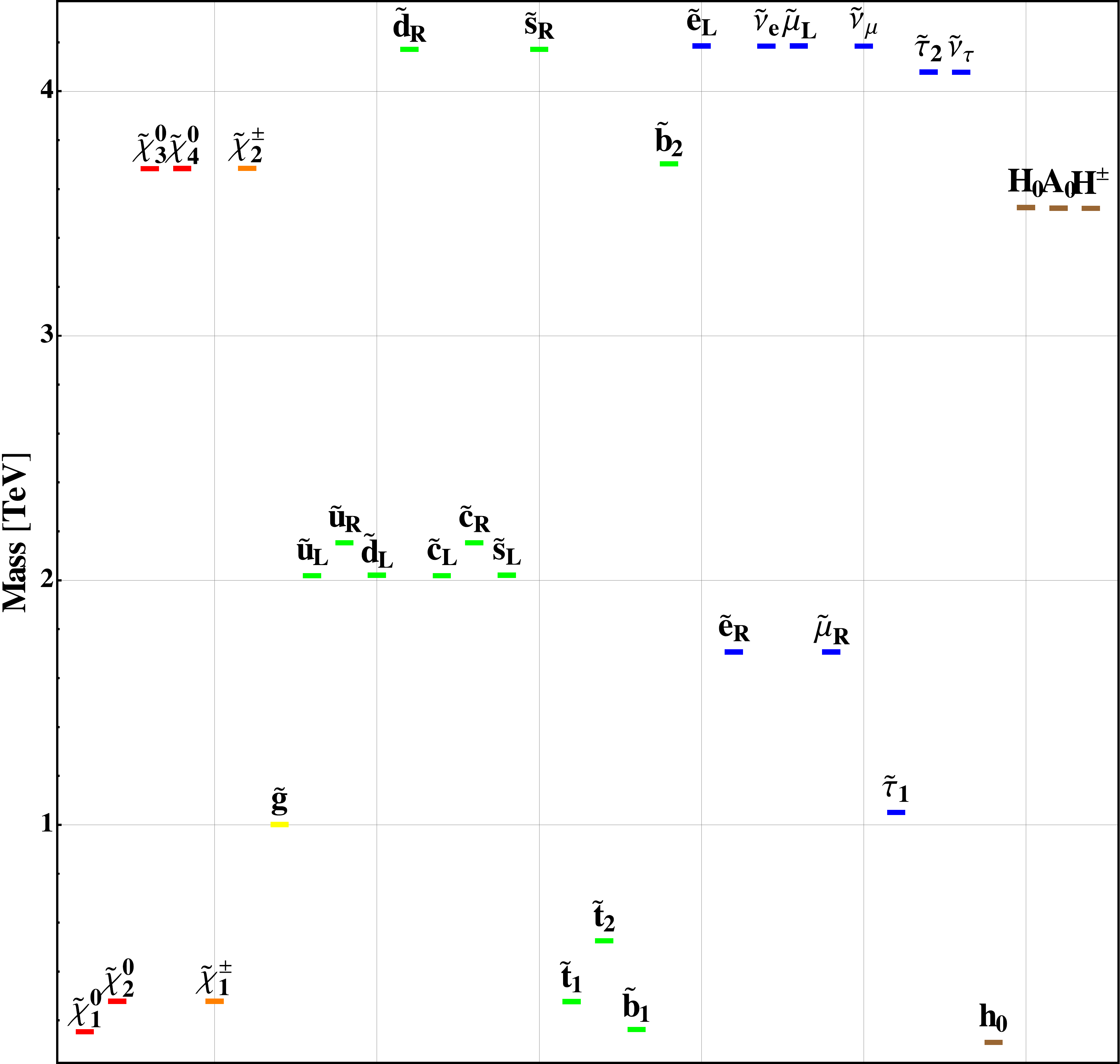}
\includegraphics[width=0.49\textwidth]{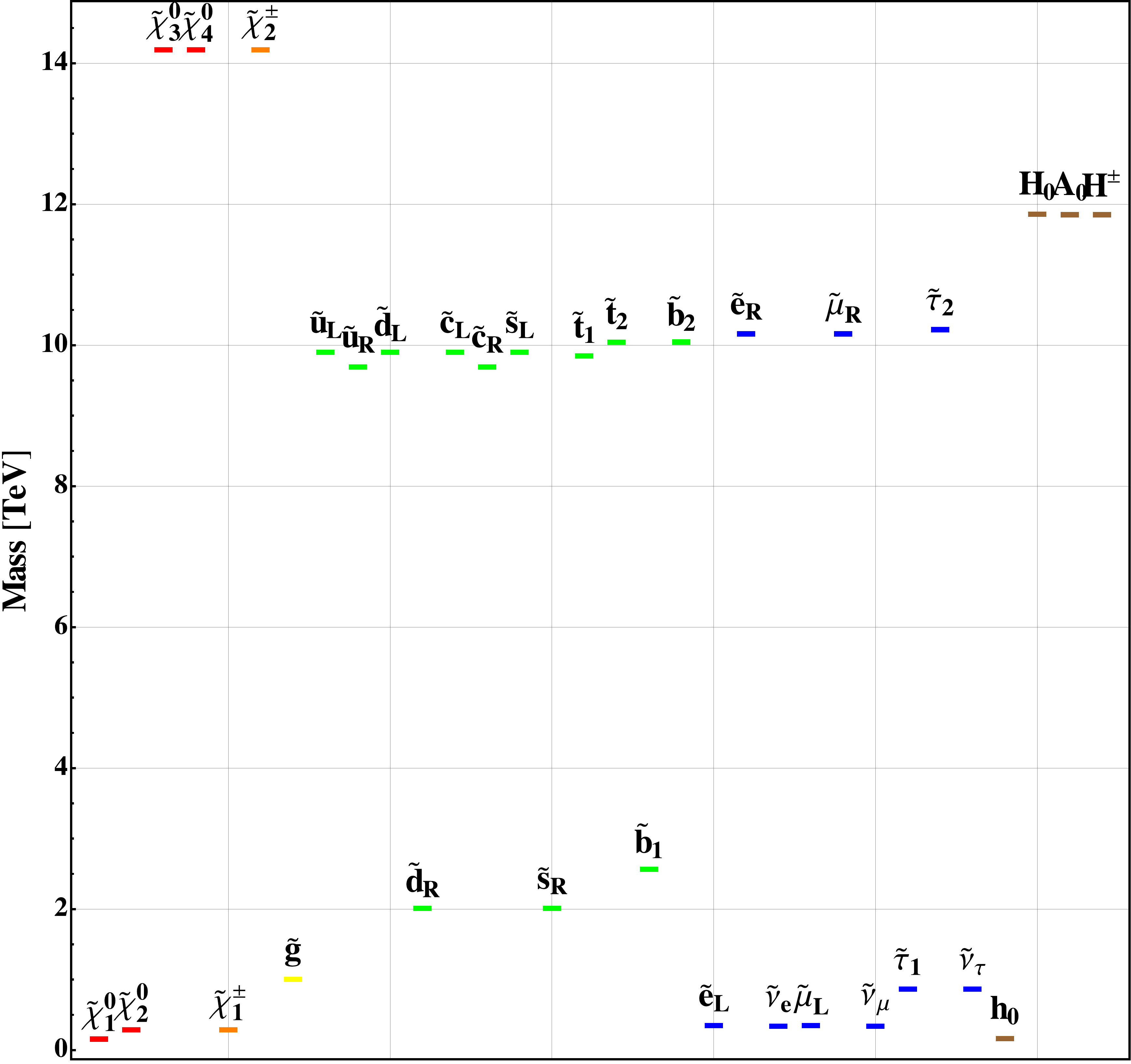}
\caption{Supersymmetric particle spectra in two example scenarios with large SO(10) D-terms based on Eq.~\eqref{eq:BenchmarkLightThird} with $m_D^2 = -(1.83 \TeV)^2$ (light third generation, left) and based on Eq.~\eqref{eq:BenchmarkLightFirst} with $m_D^2 = +(4.87 \TeV)^2$ (light first generation, right).}
\label{fig:Spectrum}
\end{figure}
We have considered three scenarios: Firstly, starting from a non-universal Higgs mass benchmark scenario, cf. Eq.~\eqref{eq:BenchmarkLightThird}, we studied the impact of the D-term $m_D^2$ on the sparticle spectrum, especially on the possibility to obtain light third generation squarks and sleptons. In particular, we found that for $m_D^2 \lesssim -m_{16_F}^2$, both stops, the lightest sbottom and the lightest stau can be very light, while the first generation squarks and sleptons are heavy. An example spectrum is shown in Figure~\ref{fig:Spectrum}~(left) for $m_D^2 \approx -(1.8 \TeV)^2 \approx - 0.5\times m_{16_F}^2$. Such a spectrum can be viable as a solution to the hierarchy problem as it keeps the fine tuning under control. It belongs to a class of Split-SUSY scenarios with a compressed spectrum \cite{Alwall:2008ve, LeCompte:2011cn, Brummer:2014yua}, with the lightest stop too light to decay into a top and the lightest neutralino. The LHC limit on the stop mass for this case is much more relaxed that in other scenarios. With a light stop mass just above the LHC limit for a compressed spectrum, $m_{\tilde t_1} \gtrsim 250 \GeV$, a rough estimate of the fine tuning would be $M_\text{SUSY}^2 / m_t^2 \approx m_{\tilde{t}_1} m_{\tilde{t}_2} / m_t^2 \approx 5$.

Secondly, we extended the previous case to make the first generation light, by way of changing the soft Higgs mass $m_{10_H}^2$. While this presents a rather extreme scenario which is fine-tuned to cancel the Yukawa contribution of the third generation states, it demonstrates the potential to deviate from the usual light stop/sbottom/stau case (although this is usually preferred due to naturalness considerations). The direct LHC limits on first and second generation slepton masses are still comparatively weak and can accommodate light sleptons $m_{\tilde{l}} \gtrsim 300 \GeV$. An example spectrum for this case is shown in Figure~\ref{fig:Spectrum}~(right) for $m_D^2 \approx +(4.9 \TeV)^2 \approx 0.3\times m_{16_F}^2$, resulting in a severely split scenario. Consequently, it requires a considerable fine-tuning, not only by manually engineering the light selectrons, but also due to the necessary cancellations of the large contributions to the Higgs mass from the heavy stops, $M_\text{SUSY}^2 /m_t^2 \approx m_{\tilde{t_1}} m_{\tilde{t}_2} / m_t^2 \approx 3 \times 10^3$. As mentioned, the main purpose of the two limiting examples provided here is to define a rough range of possible spectra in the minimal SUSY SO(10) model with large D-terms. If taken seriously, a spectrum with light first generation sleptons would naively be advantageous to explain the apparent discrepancy between the measured value of the anomalous magnetic moment of the muon $a_\mu$ and its SM prediction. Unfortunately, due to the splitting between left- and right-handed smuons in combination with the large Higgs $\mu$-term, it is not possible to appreciably raise the SUSY contribution to $a_\mu$. For $m_D^2 \gg 0$, only the right-handed down-type squarks will be light and, as we have demonstrated, this weakens the current direct LHC limit on the corresponding squark masses from $m_{\tilde q} \gtrsim 2 \TeV$ to $m_{\tilde q} \gtrsim 1 \TeV$.  

Finally, we have also briefly looked at the case of non-universal gauginos at the GUT scale. In addition to the universal case, we studied two different choices for the representation of the messenger fields; one where the messenger is in the $\mathbf{24}$ representation of the SU(5) subgroup embedded in SO(10), and one where it is in the $\mathbf{200}$ representation. The former leads to a lighter, bino-like lightest neutralino, but it negligibly affects the scalar particle masses. The latter case, leading to bino heavier than the gluino and a wino-like lightest neutralino, has a greater impact on the scalar SUSY particle masses. Both cases can of course affect the possible decay channels and therefore the visible signatures in detail. For example, raising the neutralino masses will facilitate the realization of compressed spectra and the possibility of stop-neutralino co-annihilation affecting the dark matter relic density of the universe.

\section*{Acknowledgments}
The work was supported partly by the London Centre for Terauniverse Studies (LCTS), using funding from the European Research Council via the Advanced Investigator Grant 267352. ND acknowledges support from the Alexander von Humboldt foundation. FFD and TEG would like to thank Ben Allanach, Julia Harz and Werner Porod, ND would like to thank Jad Marrouche for useful discussions.


\appendix
\section{Renormalization Group Equations}
\label{app:RGEs}
We here list the RGEs for the MSSM at one and two loop level \cite{Martin:1993zk}. In cases where the equations are analytically solvable, we provide the exact solution. Otherwise, we provide an analytical approximation. The scale parameter $t$ is defined as $t = \log\mu$ for an energy scale $\mu$.

\subsection{Gauge Couplings}
The $\beta$-functions for the gauge couplings at 1-loop are:
\begin{equation}
\frac{1}{16\pi^2}\beta_{g_a} =  \frac{dg_a}{dt} = \frac{b_a}{16\pi^2} g_a^3, \qquad (b_1,b_2,b_3) = (\tfrac{33}{5},1,-3),
\end{equation}
They are exactly solvable at 1-loop with solution ($\alpha_a = g_a^2/4\pi$):
\begin{equation}
	\alpha_a(\mu) = 	
	\frac{\alpha(M_\text{GUT})}
	{1-\tfrac{b_a}{2\pi}\alpha(M_\text{GUT})\log\tfrac{\mu}{M_\text{GUT}}}.
\end{equation}
%

\subsection{Yukawa Couplings}
Neglecting the Yukawa couplings of the first two generations, the $\beta$-functions for the 3rd generation Yukawa couplings are at 1-loop level:
\begin{align}
\notag \frac{1}{16\pi^2} \beta_{y_t} &=  \frac{dy_t}{dt} = \frac{y_t}{16\pi^2} \left(6y_t^2 + y_b^2 -\frac{16}{3}g_3^2 - 3g_2^2 - \frac{13}{15}g_1^2\right), \\
\notag \frac{1}{16\pi^2} \beta_{y_b} &= \frac{dy_b}{dt} = \frac{y_b}{16\pi^2} \left(6y_b^2 + y_t^2 + y_\tau^2 - \frac{16}{3}g_3^2 - 3 g_2^2 - \frac{7}{15}g_1^2 \right), \\ 
\frac{1}{16\pi^2} \beta_{y_\tau} &= \frac{dy_\tau}{dt} = \frac{y_\tau}{16\pi^2} \left( 4y_\tau^2 + 3 y_b^2 - 3g_2^2 - \frac{9}{5} g_1^2 \right).
\end{align}
These equations are not analytically solvable, so we will make the approximation that the $\gamma_i$'s are constant and equal to their value at the electroweak scale. The approximate solutions are therefore
\begin{align}
\notag y_t(\mu) &= \sqrt 2 \frac{m_t}{v_u} \left( \frac{\mu}{M_Z} \right)^{\gamma_t}, \\
\notag y_b(\mu) &= \sqrt 2 \frac{m_b}{v_d} \left( \frac{\mu}{M_Z} \right)^{\gamma_b},\\
y_\tau(\mu) &= \sqrt 2 \frac{m_\tau}{v_d} \left( \frac{\mu}{M_Z} \right)^{\gamma_\tau},
\end{align}
with
\begin{align}
\notag \gamma_t &= \frac{1}{16\pi^2} \left(12 \frac{m_t^2}{v_u^2} + 2 \frac{m_b^2}{v_d^2} - \frac{16}{3} g_3(M_Z)^2 - 3 g_2(M_Z)^2 - \frac{13}{15} g_1(M_Z)^2\right), \\
\notag \gamma_b &= \frac{1}{16\pi^2} \left(12 \frac{m_b^2}{v_u^2} + 2 \frac{m_t^2}{v_d^2} + 2 \frac{m_\tau}{v_d^2} - \frac{16}{3} g_3(M_Z)^2 - 3 g_2(M_Z)^2 - \frac{7}{15} g_1(M_Z)^2\right), \\
\gamma_t &= \frac{1}{16\pi^2} \left(8 \frac{m_\tau^2}{v_d^2} + 6 \frac{m_b^2}{v_d^2} - 3 g_2(M_Z)^2 - \frac{9}{5} g_1(M_Z)^2\right).
\end{align}
%

\subsection{Gaugino Masses}
The RGEs for the gauginos are very similar to the gauge couplings, and can therefore be solved analytically at 1-loop. The $\beta$-functions are
\begin{equation}
	\beta_{M_a} = 16\pi^2 \frac{dM_a}{dt} = 2 b_a g_a^2M_a,
\end{equation}
and the solution can be expressed in terms of the gauge couplings as
\begin{equation}
\label{GauginosRGEs}
	\frac{M_a(\mu)}{M_a(M_\text{GUT})} = \frac{g_a^2(\mu)}{g_a^2(M_\text{GUT})}.
\end{equation}
%

\subsection{Trilinear Couplings}
As for the Yukawa couplings, we only consider the 3rd generation trilinear couplings. Their RGEs are:
\begin{align}
\notag \frac{1}{16\pi^2} \beta_{A_t} &= \frac{dA_t}{dt} = \frac{1}{16\pi^2} \left( 12 y_t^2 A_t + 2 y_b^2 A_b + \frac{32}{3} g_3^2 M_3 + 6 g_2^2 M_2 + \frac{26}{15}g_1^2 M_1 \right), \\
\notag \frac{1}{16\pi^2} \beta_{A_b} &= \frac{dA_b}{dt} = \frac{1}{16\pi^2} \left( 12 y_b^2 A_b + 2 y_t^2 A_t + 2y_\tau^2 A_\tau + \frac{32}{3} g_3^2 M_3 + 6 g_2^2 M_2 + \frac{14}{15}g_1^2 M_1 \right), \\
 \frac{1}{16\pi^2} \beta_{A_\tau} &= \frac{dA_\tau}{dt} = \frac{1}{16\pi^2} \left( 8 y_\tau^2 A_\tau + 6 y_b^2 A_b + 6 g_2^2 M_2 + \frac{18}{5}g_1^2 M_1 \right).
\end{align}
The terms proportional to the Yukawa and trilinear couplings are not exactly solvable, thus we will make the approximation that $A_i$ is roughly constant and equal to its value at the GUT scale, $A_0$, and we solve for the Yukawa part using the approximated solution obtained above. This gives
\begin{align}
\notag A_t(\mu) &= A_0 - \frac{A_0}{8\pi^2}\left( 6 \delta_t + \delta_b)\right) - \left(\frac{16}{3} C_3^{(1)}(\mu) + 3 C_2^{(1)}(\mu) + \frac{13}{15}C_1^{(1)}(\mu) \right)m_{1/2}, \\
\notag A_b(\mu) &= A_0 - \frac{A_0}{8\pi^2}\left(\delta_t + 6 \delta_b + \delta_\tau \right) - \left( \frac{16}{3} C_3^{(1)}(\mu) + 3 C_2^{(1)}(\mu) + \frac{7}{15}C_1^{(1)}(\mu) \right)m_{1/2}, \\
A_\tau(\mu) &= A_0 - \frac{A_0}{8\pi^2}\left(4 \delta_\tau + 3 \delta_b \right) - \left( 3 C_2^{(1)}(\mu) + \frac{9}{5}C_1^{(1)}(\mu) \right) m_{1/2},
\end{align}
where $\delta_i = \frac{1}{2\gamma_i} (y_i^2 (M_\text{GUT}) - y_i^2(\mu))$ and
\begin{equation}
	C_a^{(n)}(\mu) = \frac{1}{b_a}
	\left(1 - \frac{g_a^{2n}(\mu)}{g_a^{2n}(M_\text{GUT})} \right).
\end{equation}
%

\subsection{Scalar Masses}
The $\beta$-functions for the matter sfermion masses are
\begin{align}
\notag \frac{1}{16\pi^2} \beta_{\mathbf{m}_Q^2} &= \frac{d}{dt}\mathbf{m}_Q^2 = \frac{1}{16\pi^2} \left( \mathbf{X}_u + \mathbf{X}_d - \frac{32}{3} g_3^2 M_3^2 - 6 g_2^2 M_2^2 - \frac{2}{15} g_1^2 M_1^2 + \frac{1}{5}g_1^2 S \right), \\
\notag \frac{1}{16\pi^2} \beta_{\mathbf{m}_u^2} &= \frac{d}{dt}\mathbf{m}_u^2 = \frac{1}{16\pi^2} \left( 2\mathbf{X}_u - \frac{32}{3}g_3^2M_3^2 -\frac{32}{15} g_1^2 M_1^2 - \frac{4}{5} g_1^2 S \right), \\
\notag \frac{1}{16\pi^2} \beta_{\mathbf{m}_d^2} &= \frac{d}{dt}\mathbf{m}_d^2 = \frac{1}{16\pi^2} \left( 2\mathbf{X}_d - \frac{32}{3}g_3^2 M_3^2 - \frac{8}{15} g_1^2 M_1^2 + \frac{2}{3} g_1^2 S \right), \\
\notag \frac{1}{16\pi^2} \beta_{\mathbf{m}_L^2} &= \frac{d}{dt}\mathbf{m}_L^2 = \frac{1}{16\pi^2} \left( \mathbf{X}_e - 6g_2^2 M_2^2 - \frac{6}{5} g_1^2 M_1^2 - \frac{3}{5} g_1^2 S \right), \\
\frac{1}{16\pi^2} \beta_{\mathbf{m}_e^2} &= \frac{d}{dt}\mathbf{m}_e^2 = \frac{1}{16\pi^2} \left( 2\mathbf{X}_e - \frac{24}{5} g_1^2 M_1^2 + \frac{6}{5} g_1 S \right),
\end{align}
where $\mathbf{X}_i$ are $3\times 3$ matrices proportional to the $3\times 3$ Yukawa matrices. and 
\begin{equation}
	S = m_{H_u}^2 - m_{H_d}^2 + \text{Tr} \left( \mathbf{m}_Q^2 - 
	\mathbf{m}_L^2 - 2\mathbf{m}_u^2 + \mathbf{m}_d^2 + \mathbf{m}_e^2 \right).
\end{equation}
Neglecting the Yukawa couplings for the first and second generations, the (3,3) components of the $\mathbf{X}_i$ can be written as
\begin{align}
  X_t &= 
  2y_t^2 \left(m_{H_u}^2 + m_{Q_3}^2 + m_{u_3}^2 + A_t^2 \right), \notag \\
  X_b &= 
  2y_b^2 \left(m_{H_d}^2 + m_{Q_3}^2 + m_{d_3}^2 + A_b^2 \right), \notag \\
  X_\tau &= 
  2y_\tau^2 \left(m_{H_d}^2 + m_{L_3}^2 + m_{e_3}^2 + A_\tau^2 \right).
\end{align}
The gauge components are exactly solvable, as is the dependence on $S$. Hence, for the first two generations it is possible to arrive at an exact analytical solution at 1-loop:
\begin{align}
\notag m_{Q_{1,2}}^2 &= m_{16_F}^2 + \left(1 + \frac{2}{5}C_1^{(1)}\right)m_D^2 + \left(\frac{8}{3}C_3^{(2)} + \frac{3}{2}C_2^{(2)} + \frac{1}{30}C_1^{(2)}\right) m_{1/2}^2, \\
\notag m_{u_{1,2}}^2 &= m_{16_F}^2 + \left(1 - \frac{8}{5}C_1^{(1)}\right)m_D^2 + \left(\frac{8}{3}C_3^{(2)} + \frac{8}{15}C_1^{(2)}\right) m_{1/2}^2, \\
\notag m_{d_{1,2}}^2 &= m_{16_F}^2 + \left(-3 + \frac{4}{5}C_1^{(1)}\right)m_D^2 + \left(\frac{8}{3}C_3^{(2)} + \frac{2}{15}C_1^{(2)}\right) m_{1/2}^2, \\
\notag m_{L_{1,2}}^2 &= m_{16_F}^2 + \left(-3 - \frac{6}{5}C_1^{(1)}\right)m_D^2 + \left(\frac{3}{2}C_2^{(2)} + \frac{3}{10}C_1^{(2)}\right) m_{1/2}^2, \\
m_{e_{1,2}}^2 &= m_{16_F}^2 + \left(1 + \frac{12}{5}C_1^{(1)}\right)m_D^2 + \frac{6}{5}C_1^{(2)} m_{1/2}^2.
\end{align}
The third sfermion generations have an extra dependence on the Yukawa and trilinear couplings via the terms $X_t, X_b, X_\tau$. Their RGEs cannot be solved analytically. We approximate the dependence on the scalar masses and trilinear couplings by taking them constant with values given by the geometrical average of their values at the GUT scale and the SUSY scale. Using the approximate solution for the Yukawa couplings, this gives
\begin{align}
  m^2_{Q_3}       &= m^2_{Q_{1,2}} -\Delta_t - \Delta_b, \notag \\
  m^2_{\bar{u}_3} &= m^2_{\bar{u}_{1,2}}  - 2\Delta_t, \notag \\
  m^2_{\bar{d}_3} &= m^2_{\bar{d}_{1,2}} - 2\Delta_b, \notag \\
  m^2_{L_3}       &= m^2_{L_{1,2}} - \Delta_\tau, \notag \\
  m^2_{\bar{e}_3} &= m^2_{\bar{e}_{1,2}} - 2\Delta_\tau,
\end{align}
with
\begin{align}
  \Delta_t &= 
  \frac{1}{8\pi^2}\delta_t 
  \left( m_{10_H}^2 + 2|m_{16_F}| \tilde{M} 
  + A_0 A_t(\tilde{M}) \right), \notag \\
  \Delta_b &= 
  \frac{1}{8\pi^2}\delta_b 
  \left( m_{10_H}^2 + 2|m_{16_F}| \tilde{M} 
  + A_0 A_b(\tilde{M})\right), \notag \\
  \Delta_\tau &= 
  \frac{1}{8\pi^2}\delta_\tau 
  \left( m_{10_H}^2 + 2|m_{16_F}| \tilde{M} 
  + A_0 A_\tau(\tilde{M}) \right).
\end{align}
Finally, the Higgs doublet soft masses have similar RGEs to the other scalars,
\begin{align}
  \frac{1}{16\pi^2} \beta_{m_{H_u}^2} &= \frac{d}{dt} m_{H_u}^2 =  
  \frac{1}{16\pi^2} \left( 3X_t - 6g_2^2 M_2^2 - \frac{6}{5} g_1^2 M_1^2 
  + \frac{3}{5}  g_1^2 S \right), \notag \\
  \frac{1}{16\pi^2}\beta_{m_{H_d}} &= \frac{d}{dt}m_{H_d} = 
  \frac{1}{16\pi^2} \left( 3X_b + X_\tau - 6 g_2^2 M_2^2 
  - \frac{6}{5} g_1^2 M_1^2 - \frac{3}{5} g_1^2 S \right),
\end{align}
and can be solved using the same approximation yielding 
\begin{align}
  m^2_{H_u} &= 
  m^2_{10_H} + \left(- 2 + \frac{6}{5}C_1^{(1)} \right) m_D^2 
  + \left( \frac{3}{2}C_2^{(2)} + \frac{3}{10}C_1^{(2)} \right) m_{1/2}^2 
  - 3 \Delta_t, \notag \\
  m^2_{H_d} &= 
  m^2_{10_H} + \left( 2 - \frac{6}{5}C_1^{(1)} \right) m_D^2 
  + \left( \frac{3}{2}C_2^{(2)} + \frac{3}{10}C_1^{(2)} \right) m_{1/2}^2 
  - 3 \Delta_b - \Delta_\tau.
\end{align}

\subsection{$\mu_H$ and $B$ Terms}
Both $\mu_H$ and $B$ can be fixed at the electroweak scale by requiring successful electroweak symmetry breaking. Therefore we will use the electroweak scale ($M_Z$) as the reference point to solve the RGEs. The RGEs are
\begin{align}
  \frac{1}{16\pi^2} \beta_{\mu_H} &= \frac{d\mu_H}{dt} = 
  \frac{\mu_H}{16\pi^2} 
  \left( 3 y_t^2 + 3 y_b^2 + y_\tau^2 - 3g_2^2 
  - \frac{3}{5}g_1^2 \right), \notag \\
  \frac{1}{16\pi^2} \beta_B &= \frac{dB}{dt} = 
  \frac{1}{16\pi^2} \left( 3 A_t y_t^2 + 6 A_b y_b^2 
  + 2A_\tau y_\tau^2 + 6g_2^2 M_2 + \frac{6}{5} g_1^2 M_1 \right).
\end{align}
The solution is calculated using the analogous approximations we used for the Yukawa couplings and the trilinear terms, respectively,
\begin{align}
  \mu_H(\mu) &= 
  \mu_H(M_Z) \left( \frac{\mu}{M_Z} \right)^{\gamma_{\mu_H}}, \notag \\
  B(\mu) &= 
  B(M_Z) - \frac{A_0}{8\pi^2} ( 6 \delta'_t+ 6 \delta'_b + \delta'_\tau ) 
  - \left( 3 {C'}_2^{(1)}- \frac{3}{5}{C'}_1^{(1)} \right) m_{1/2},
\end{align}
where $\delta'_i$ and ${C'}_a^{(1)}$ are the same as $\delta_i$ and $C_a^{(1)}$ defined before but with $M_Z$ as a reference scale instead of $M_\text{GUT}$, and
\begin{equation}
  \gamma_{\mu_H} = \frac{1}{16\pi^2} \left( 6 \frac{m_t^2}{v_u^2} 
  + 6 \frac{m_b^2}{v_d^2} + 2\frac{m_\tau^2}{v_d^2} - 3 g_2(M_Z)^2 
  - \frac{3}{5} g_1(M_Z)^2 \right).
\end{equation}
The values of $\mu_H$ and $B$ at the EW scale are given at 1-loop by ($t_\beta \equiv \tan\beta$) \cite{Drees:2004jm}
\begin{align}
	\mu_{H,\text{tree}}^{2} &= 
	-\frac{m_{H_d}^2}{1 - t_\beta^2} - \frac{m_{H_u}^2}{1 - t_\beta^{-2}} 
	- \frac{1}{8} (g_2^2 + \frac{3}{5} g_1^2) (v_d^2+v_u^2), \notag\\
	B_\text{tree}^{2} &=  
	\frac{1}{\mu_{H,tree}} \left(\frac{m_{H_d}^2 
	- m_{H_u}^2}{t_\beta - t_\beta^{-1}} 
	- \frac{1}{4} (g_2^2 + \tfrac{3}{5}g_1^2) v_u v_d \right), \notag\\
	\mu_H^2(M_Z) &= 
	\mu_{H,\text{tree}}^{2} - \frac{3y_t^2}{32 \pi^2 (1-t_\beta^2)} 
	\frac{\mu_{H,\text{tree}} (\mu_{H,\text{tree}}
	- A_t)t_\beta}{m_{\tilde{t}_1}^2 
	- m_{\tilde{t}_2}^2}\left(f(m_{\tilde{t}_1}^2) 
	- f(m_{\tilde{t}_2}^2)\right) 
	+ \frac{3y_t^2}{32 \pi^2} \frac{t_\beta^2}{1-t_\beta^{-2}}\notag\\
	& 
	\times\left(f(m_{\tilde{t}_1}^2) + f(m_{\tilde{t}_2}^2) 
	- 2 f(m_t^2) + \frac{A_t (A_t - \mu_{H,\text{tree}}
	t_\beta^{-1})}{m_{\tilde{t}_1}^2 - m_{\tilde{t}_2}^2}
	\left(f(m_{\tilde{t}_1}^2) - f(m_{\tilde{t}_2}^2)\right)\right),
\end{align}
with the stop mass eigenvalues $m_{\tilde{t}_{1,2}}$ and the function
\begin{align}
	f(x) = 2 x \left(\log\frac{x}{M_\text{SUSY}^2} -1\right).
\end{align}
%
\subsection{Two-Loop Corrections}
We employ two loop corrections only for the scalar masses, because for large $m_D^2$ and consequently large $m_{16_F}$, their contribution can be sizable. The relevant 2-loop beta functions, in which we neglect the Yukawa couplings of the first two generations, are given by \cite{Martin:1993zk}
\begin{align}
\notag \beta_{m_Q^2}^{(2)} =  &- 20(m_Q^2 + m_{H_u}^2 + m_u^2)y_u^4 - 20(m_Q^2 +  m_{H_d}^2 + m_d^2) y_d^4  \\
\notag & - 2(m_Q^2 + m_L^2 + 2m_{H_d}^2 + m_d^2 + m_e^2) y_d^2y_e^2 \\
\notag &- 40 A_t^2 y_u^2 - 40 A_b^2 y_d^2 - 2 y_d^2 y_e^2(A_b + A_\tau)^2 \\
\notag & + \frac{2}{5} g_1^2 \left\{ 4(m_Q^2 + m_{H_u}^2 + m_u^2 + A_t^2 -  (M_1 + M_1^*) A_t + 2 |M_1|^2)y_u^2 \right.\\
\notag & + \left. 2(m_Q^2 + m_{H_d}^2 + m_d^2 + A_b^2 - (M_1+M_1^*) A_b + 2 |M_1|^2)y_d^2 \right\} \\
\notag &- \frac{128}{3} g_3^4 |M_3|^2 + 32 g_3^2 g_2^2(|M_3|^2 + |M_2|^2 + \Re[M_2M_3^*]) \\
\notag&+ \frac{32}{45} g_3^2g_1^2(|M_3|^2 + |M_1|^2 + \Re[M_1M_3^*]) + 33 g_2^4 |M_2|^2 \\
\notag &+ \frac{2}{5} g_2^2 g_1^2 (|M_2|^2 + |M_1|^2 + \Re[M_1M_2^*]) + \frac{199}{75}g_1^4|M_1|^2 + \frac{16}{3} g_3^2 \sigma_3 \\
&+ 3 g_2^2 \sigma_2 + \frac{1}{15} g_1^2 \sigma_1 + \frac{2}{5} g_1^2 S',
\end{align}
\begin{align}
\notag \beta_{m_L^2}^{(2)} = &- 12 (m_L^2 + m_{H_d}^2 + m_e^2) y_e^4 \\
\notag &- 6(m_Q^2 + m_L^2 + 2m_{H_d}^2 + m_d^2 + m_u^2) y_d^2 y_e^2 \\
\notag &- 24 A_\tau^2 y_e^4 - 6y_d^2 y_e^2 (A_b + A_\tau)^2 \\
\notag & + \frac{12}{5} g_1^2 \left\{ m_L^2 + m_{H_d}^2 + m_e^2 + A_\tau^2 - (M_1+M_1^*) A_\tau + 2 |M_1|^2 \right\} y_e^2 \\
\notag & + 33 g_2^4 |M_2|^2 + \frac{18}{5} g_2^2 g_1^2 (|M_2|^2 + |M_1|^2 + \Re[M_1M_2^*]) + \frac{621}{25}g_1^4|M_1|^2 \\
&+ 3 g_2^2 \sigma_2 + \frac{3}{5} g_1^2 \sigma_1 - \frac{6}{5} g_1^2 S',
\end{align}
\begin{align}
\notag \beta_{m_u^2}^{(2)} =  &- 32 (m_u^2 + m_{H_u}^2 + m_Q^2)y_u^4 \\
\notag &- 4 (m_u^2 + 2m_Q^2 + m_d^2 + m_{H_u}^2 + m_{H_d}^2) y_u^2 y_d^2 \\
\notag &- 64 A_t^2 y_u^4 - 4 y_u^2y_d^2(A_t + A_b)^2 \\
\notag &+ \left[12 g_2^2 - \frac{4}{5} g_1^2\right] \left\{ m_u^2 + m_Q^2 + m_{H_u}^2 + A_t^2\right\}y_u^2 \\
\notag & + 12 g_2^2 \left\{ 2|M_2|^2 - (M_2 + M_2^*)A_t\right\} y_u^2 - \frac{4}{5} \left\{ 2|M_1|^2 - (M_1+M_1^*) A_t\right\} y_u^2 \\
\notag &- \frac{128}{3} g_3^4 |M_3|^2 + \frac{512}{45} g_3^2g_1^2(|M_3|^2 + |M_1|^2 + \Re[M_1M_3^*]) + \frac{3424}{75}g_1^4|M_1|^2 \\
&+ \frac{16}{3} g_3^2 \sigma_3 +\frac{16}{15} g_1^2 \sigma_1 - \frac{8}{5} g_1^2 S',
\end{align}
\begin{align}
\notag \beta_{m_d^2}^{(2)} =  &- 32 (m_d^2 + m_{H_d}^2 + m_Q^2) y_d^4 \\
\notag &- 4(m_d^2 + 2m_Q^2 + m_u^2 + m_{H_u}^2 + m_{H_d}^2) y_u^2 y_d^2  - 8 (m_L^2 + m_e^2)y_d^2 y_e^2 \\
\notag & - 64 A_b^2 y_d^4 - 4 y_d^2 y_u^2 (A_t + A_b)^2 - 4 y_d^2 y_e^2 (A_b + A_\tau)^2 \\
\notag  & + \left[12 g_2^2 + \frac{4}{5}g_1^2 \right] \left\{ m_d^2 + m_{H_d}^2 + m_Q^2 + A_b^2\right\} y_d^2 \\
\notag & + 12 g_2^2 \left\{2 |M_2|^2 - (M_2+M_2^*) A_b \right\} y_d^2 + \frac{4}{5}g_1^2 \left\{ 2|M_1|^2 - (M_1 + M_1^*) A_b \right\} y_d^2 \\
\notag&- \frac{128}{3} g_3^4 |M_3|^2 +  \frac{128}{45} g_3^2g_1^2(|M_3|^2 + |M_1|^2 + \Re[M_1M_3^*]) + \frac{808}{75}g_1^4|M_1|^2 \\
&+ \frac{16}{3} g_3^2 \sigma_3 + \frac{4}{15} g_1^2 \sigma_1 + \frac{4}{5} g_1^2 S',
\end{align}
\begin{align}
\notag \beta_{m_e^2}^{(2)} =  &- 16 (m_e^2 + m_{H_d}^2 + m_L^2) y_e^2  - 12(m_e^2 + m_L^2 + m_d^2 + 2m_{H_d}^2)y_d^2 y_e^2 \\
\notag &- 32 A_\tau y_e^4 - 4 y_e^2 y_d^2 (A_\tau + A_d)^2 + (12 g_2^2 + \frac{12}{5} g_1^2 )\left\{ m_e^2 + m_L^2 + m_{H_d}^2 + A_\tau^2 \right\} y_e^2 \\
\notag&+ 12 g_2^2 \left\{2 |M_2|^2 - (M_2 + M_2^*) A_\tau \right\} y_e^2 - \frac{12}{5} \left\{ 2|M_1|^2 - (M_1+M_1^*) A_\tau \right\} y_e^2 \\ 
&+ \frac{2808}{25}g1_4|M_1|^2 + \frac{12}{5}g_1^2\sigma_1 + \frac{12}{5}g_1^2 S',
\end{align}
\begin{align}
\notag \beta_{m_{H_u}^2}^{(2)} = &- 36 (m_{H_u}^2 + m_Q^2 + m_u^2) y_u^4 - 6 (m_{H_u}^2 + m_{H_d}^2 + 2m_Q^2 + m_u^2 + m_d^2) y_u^2y_d^2 \\
\notag &- 72 A_t^2 y_u^4 - 6 y_u^2 y_d^2 (A_u + A_d)^2  + \left[32 g_3^2 + \frac{8}{5} g_1^2\right] \left\{ m_{H_u}^2 + m_Q^2 + m_u^2 + A_t^2 \right\} y_u^2 \\
\notag & + 32 g_3^2 \left\{ 2 |M_3|^2 - (M_3 + M_3^*) A_t \right\} y_u^2  + \frac{8}{5} g_1^2 \left\{ 2 |M_1|^2 - (M_1 + M_1^*)A_t \right\} y_u^2 \\
\notag &+ 33 g_2^4|M_2|^2 + \frac{18}{5}g_2^2g_1^2(|M_2|^2 + |M_1|^2 + \Re[M_1M_2^*]) + \frac{621}{25}g_1^4|M_1|^2 \\
&+ 3g_2^2 \sigma_2 + \frac{3}{5}g_1^2\sigma_1 -+\frac{6}{5}g_1^2 S',
\end{align}
\begin{align}
\notag\beta_{m_{H_d}^2}^{(2)} = & - 36 (m_{H_d}^2  + m_Q^2 + m_d^2 ) y_d^4 - 6 (m_{H_u}^2 + m_{H_d}^2 + 2m_Q^2 + m_u^2 + m_d^2) y_u^2y_d^2 \\
\notag & - 12 (m_{H_d}^2 + m_L^2 + m_e^2) y_e^4 - 72 A_b^2 y_d^4  - 6 y_u^2 y_d^2(A_t + A_b)^2 - 24 A_\tau^2 y_e^4 \\
\notag & + \left[33 g_3^2 - \frac{4}{5}g_1^2\right] \left\{m_{H_d}^2 + m_Q^2 + m_d^2 + A_b^2\right\} y_d^2 \\
\notag &+ 32 g_3^2 \left\{ 2|M_3|^2 - (M_3 + M_3^*)A_b^2\right\} y_d^2- \frac{4}{5} g_1^2 \left\{ 2|M_1|^2 - (M_1 + M_1^*) A_b^2 \right\} y_d^2 \\
\notag &+ \frac{12}{5}g_1^2 \left\{m_{H_d}^2 + m_L^2 + m_e^2 + A_\tau^2 + 2|M_1|^2 - (M_1 + M_1^*)A_\tau\right\} y_e^2 \\
\notag & + 33 g_2^4|M_2|^2 + \frac{18}{5}g_2^2g_1^2(|M_2|^2 + |M_1|^2 + \Re[M_1M_2^*]) + \frac{621}{25}g_1^4|M_1|^2 \\
&+ 3g_2^2 \sigma_2 + \frac{3}{5}g_1^2\sigma_1 - \frac{6}{5}g_1^2 S',
\end{align}
In the above equations, the following definitions apply:
\begin{align}
\notag S' &= - (3 m_{H_u}^2 + m_Q^2 - 4m_u^2) y_u^2 + (3 m_{H_d}^2 - m_Q^2 - 2 m_d^2 ) y_d^2 + (m_{H_d}^2 + m_L^2 - m_e^2) y_e^2\\
\notag &+ \left[\frac{3}{2} g_2^2 + \frac{3}{10} g_1^2 \right] \left\{ m_{H_u}^2 - m_{H_d}^2 + \Tr(m_L^2)\right\} + \left[\frac{8}{3} g_3^2 + \frac{3}{2} g_2^2 + \frac{1}{30}g_1^2 \right]\Tr(m_Q^2)\\
&- \left[ \frac{16}{3}g_3^2 + \frac{16}{15} g_1^2\right] \Tr(m_u^2) + \left[\frac{8}{3}g_3^2 + \frac{2}{15}g_1^2\right]\Tr(m_d^2) + \frac{6}{5}g_1^2 \Tr(m_e^2),\notag\\
\notag \sigma_1 &= \frac{1}{5} g_1^2  \left\{ 3(m_{H_u}^2 + m_{H_d}^2 ) + \Tr(m_Q^2 + 3m_L^2 + 8m_u^2 + 2m_d^2 + 6m_e^2)\right\},\\
\notag \sigma_2 &= g_2^2 \left\{ m_{H_u}^2 + m_{H_d}^2 + \Tr(3m_Q^2 + m_L^2) \right\}, \\
\sigma_3 &= g_3^2 \Tr(2m_Q^2 + m_u^2 + m_d^2).
\end{align}
The approximate solutions of this second loop correction are obtained by taking the value of the beta functions as constant and equal to the values at the GUT scale, using the 1-loop solutions, and integrating over scales,
\begin{equation}
  m_{i,\text{2-loop}}^2 = 
  m_{i,\text{1-loop}}^2 - \frac{1}{(16 \pi^2)^2} 
  \beta_{m_{i,\text{1-loop}}^2}^{(2)} (M_\text{GUT}) \log\frac{M_\text{GUT}}{M_\text{SUSY}}.
\end{equation}
%

\bibliographystyle{h-physrev4}
\bibliography{references}

\end{document}